\documentclass[11pt]{article}
\pdfoutput=1

\usepackage{jheppub}

\usepackage[svgnames]{xcolor}
\usepackage{float}
\usepackage[utf8]{inputenc}
\usepackage{slashed,verbatim}
\pdfoutput=1
\usepackage[most]{tcolorbox}
\usepackage{epigraph,lipsum}
\usepackage{fancyhdr}
\usepackage{epigraph}
\usepackage{graphicx}

\usepackage[normalem]{ulem}

\makeatletter
\def\@fpheader{\relax}
\makeatother

\usepackage{framed}
\usepackage{caption}

\usepackage{lipsum}

\usepackage[framemethod=tikz]{mdframed}

\usepackage{amsmath,epsfig}
\usepackage{amssymb,amsfonts}
\usepackage{latexsym}
\usepackage{graphicx}

\usepackage{subeqnarray}

\usepackage{graphicx}

\def\be{\begin{equation}}
\def\ee{\end{equation}}
\def\bea{\begin{eqnarray}}
\def\eea{\end{eqnarray}}

\newcommand{\bear}{\begin{eqnarray}}

\newcommand{\eear}{\end{eqnarray}}

\newcommand{\bsea}{\begin{subeqnarray}}
\newcommand{\esea}{\end{subeqnarray}}
\newbox\pippobox

\def\cf{c.f.~}
\def\ie{i.e.~}
\def\eg{e.g.~}

\def\6{\partial}

\newcommand{\comments}[1]{}
\allowdisplaybreaks[3]

\preprint{IFT-UAM/CSIC-19-54}

\begin{document}


\title{\centering \Huge Holographic plasmon relaxation with and without broken translations}
\author[a]{Matteo Baggioli}
\author[b]{, Ulf Gran}
\author[a]{, Amadeo Jimenez Alba}
\author[b]{, Marcus Torns\"{o}}
\author[c]{, Tobias Zingg}
\vspace{0.1cm}
\affiliation[a]{Instituto de Fisica Teorica UAM/CSIC,
c/ Nicolas Cabrera 13-15, Cantoblanco, 28049 Madrid, Spain}
\affiliation[b]{Department of Physics, Division for Theoretical Physics, Chalmers University of Technology
SE-412 96 G\"{o}teborg, Sweden}
\affiliation[c]{Nordita, Stockholm University and KTH Royal Institute of Technology
Roslagstullsbacken 23, SE-106 91 Stockholm, Sweden}

\emailAdd{matteo.baggioli@uam.es}
\emailAdd{ulf.gran@chalmers.se}
\emailAdd{amadeo.jimenez@gmail.com}
\emailAdd{marcus.tornso@chalmers.se}
\emailAdd{zingg@nordita.org}

\vspace{1cm}

\abstract{We study the dynamics and the relaxation of bulk plasmons in strongly coupled and quantum critical systems using the holographic framework. We analyze the dispersion relation of the plasmonic modes in detail for an illustrative class of holographic bottom-up models. Comparing to a simple hydrodynamic formula, we entangle the complicated interplay between the three least damped modes and shed light on the underlying physical processes. Such as the dependence of the plasma frequency and the effective relaxation time in terms of the electromagnetic coupling, the charge and the temperature of the system. Introducing momentum dissipation, we then identify its additional contribution to the damping. Finally, we consider the spontaneous symmetry breaking (SSB) of translational invariance. Upon dialing the strength of the SSB, we observe an increase of the longitudinal sound speed controlled by the elastic moduli and a decrease in the plasma frequency of the gapped plasmon. We comment on the condensed matter interpretation of this mechanism.
}

\maketitle

\section{Introduction}

\begin{figure}[t!]
    \centering
   \tcbox[colframe=green!30!black,
           colback=white!30]{ \includegraphics[width=5cm]{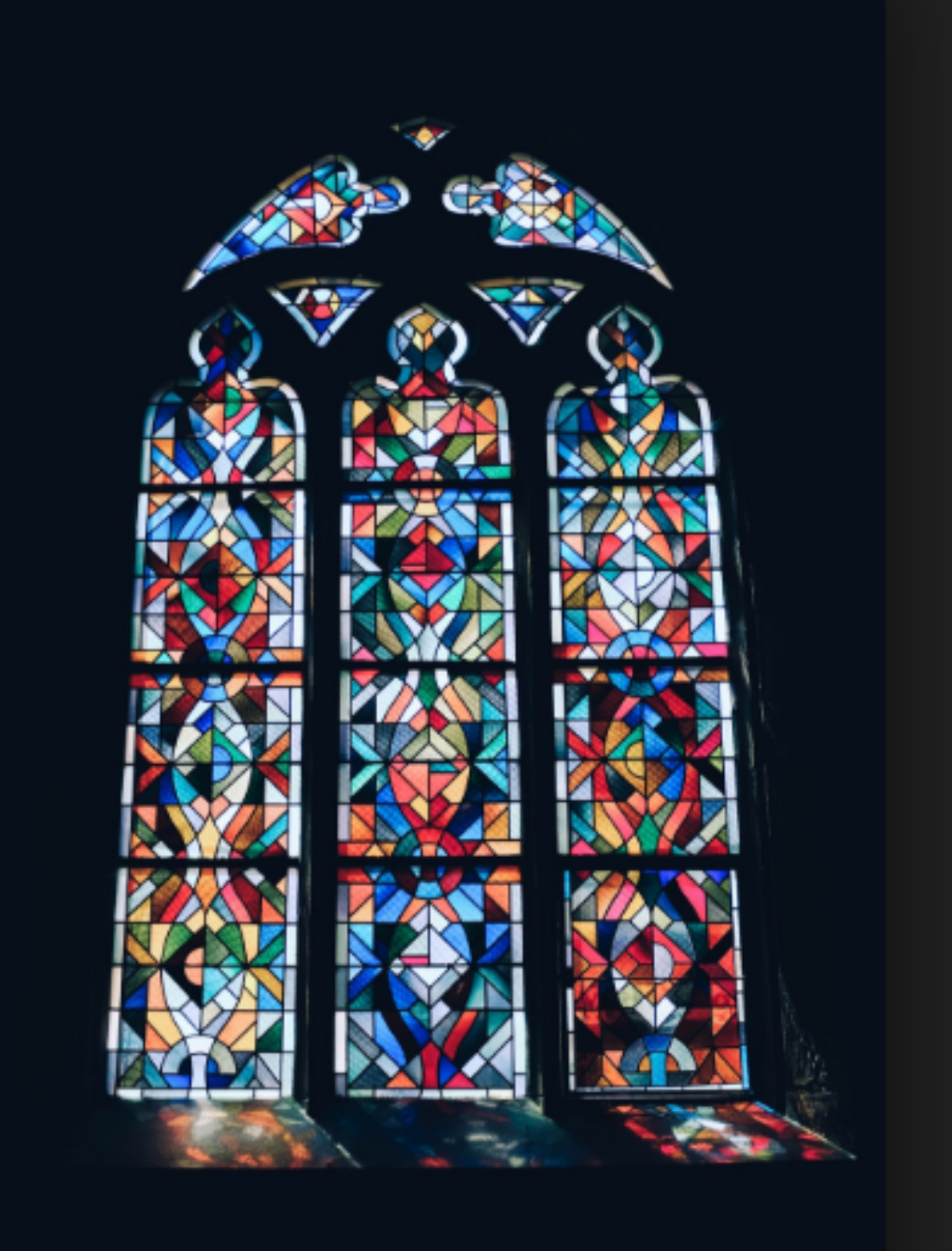}
    \quad
    \includegraphics[width=8.7cm]{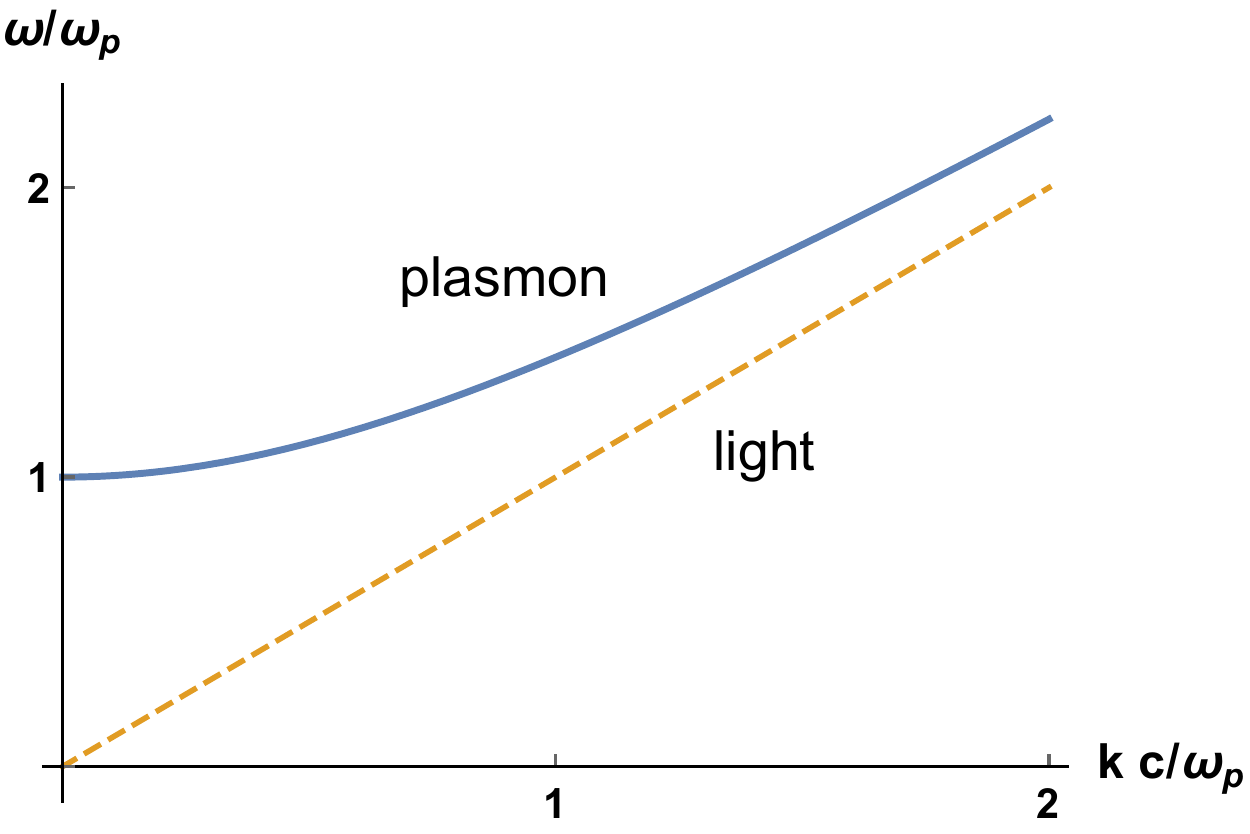}}
    \caption{\textbf{Left: } One example of the applications of plasmons in art in the form of a stained glass window in Notre Dame, Paris. \textbf{Right: } The typical bulk plasmon dispersion relation $\omega^2=\omega_p^2\,+\,c^2\,k^2$ compared with the photon dispersion relation in vacuum.}
    \label{fig0}
\end{figure}

Plasma oscillations, whose quanta are called plasmons, are collective electronic oscillations in metals. Their history is an incredible and successful connubium between art and technology. Already in the 4th century, Romans manufactured dichroic glass, used for example in the Lycurgus Cup, where plasmon effects from gold and silver particles dispersed in the glass matrix interplay with transmitted and reflected light at certain wavelengths. These techniques in staining glass where later refined and became one of the major artistic techniques in the Middle Ages~\cite{schedula}, where stunning artworks with stained glass were created, \cf figure~\ref{fig0}.
Although plasmonic effects have been known for over a millennium, a complete understanding of the underlying physics of these phenomena was not accomplished until the 1970s, which marks the beginning of modern plasmon-based applications. The technological uses of plasmons have been motivated by the attempt to overcome the diffraction limit of light and by their ability to highly enhance the electric field intensity. The modern applications are limitless, from solar cell  and cosmetics to a completely new branch of science known as plasmonics~\cite{szunerits2015introduction, Maier:2007:10.1007/0-387-37825-1}.

Plasmons appear as a result of interaction processes between electromagnetic radiation and conduction electrons and their main features can be understood with classical electromagnetism~\cite{jackson1975classical}. The dispersion relation of the longitudinal plasmon modes is characterized by the longitudinal part of the (tensorial) dielectric function $\epsilon_{ij} = \frac{\partial D_i}{\partial E_j}$ -- see e.g.~\cite{nozieres1999theory} for details. More precisely, the condition that a plasmon excitation leads to an oscillation in polarization $P$ without the influence of a change in external field $D$ leads to the condition
\begin{equation}
    \epsilon_L(k,\omega)\,=\,0\label{rel0}\,,
\end{equation}
which we therefore will call \emph{`plasmon condition'} in the following.
Using Maxwell's equations in a medium and basic definitions, the dielectric function can also be shown to be related to other elementary properties, like the conductivity of the material
\begin{equation}
    \epsilon_L(k,\omega)\,=\,1\,+\,\frac{i}{\omega}\,\frac{4\,\pi\,\sigma_L(k,\omega)}{\epsilon_0}\label{rel}\,,
\end{equation}
where $\epsilon_0$ is the vacuum dielectric constant and $\sigma_L(k,\omega)$ the spatially resolved longitudinal conductivity.
Over a wide range of frequencies, the optical properties of metals can be explained using the \textit{free electrons gas} approximation. This model idealizes the physics of a metal considering it as a gas of free electrons with number density
$n$ moving against a fixed background of positive ion cores. The electrons are driven by an applied external electromagnetic
field, and their motion is damped with a characteristic collision frequency $\gamma=1/\tau$ . The timescale $\tau$ is labelled as the relaxation time of the free
electron gas. 
For a free electron gas it is then a standard calculation to derive the dielectric function\footnote{Here, for simplicity, we omit the sub-index $L$ indicating that the dielectric function we are after relates to the longitudinal collective modes of the system.}~\cite{nozieres1999theory} ,
\begin{equation}
    \epsilon(\omega)\,=\,1\,-\,\frac{\omega_p^2}{\omega^2\,+\,i\,\gamma\,\omega}\,,
\end{equation}
where, crucially, we define the quantity
\begin{equation}
    \omega_p^2\,=\,\frac{4\,\pi\,n\,e^2}{m^*\,\epsilon_0}\,,
\end{equation}
as the plasma frequency of the metal. Here we use the symbol $m^*$ to denote the effective mass of the electronic quasi-particles, which can differ significantly from the microscopic electron mass\footnote{Let us already mention that no quasi-particles are present in the holographic picture we discuss in this paper. Therefore, this weakly coupled logic is a good guidance but it should not be taken too seriously.}.\\
For large frequencies $\omega \tau \gg 1$, we can neglect the damping effects and the dielectric function is purely real:
\begin{equation}
    \epsilon(\omega)\,=\,1\,-\,\frac{\omega_p^2}{\omega^2}\,,
\end{equation}
\ie the known result for the underdamped free electron plasma (without taking into account interband transitions). In the other limit, $\omega \tau  \ll 1$, the imaginary part of the dielectric function cannot be discarded and the physics is very different. In this last regime, the metals are mainly absorbing and the fields fall off inside the sample following  Beer's law $ \sim\, e^{-z/\delta}$,
where $\delta$ is known as the skin-depth. As we will see later, these features can be linked to the so-called $k$-gap dispersion relation~\cite{Baggioli:2019jcm} which will appear in our model as well.

The nature of plasmons in ordinary, and weakly coupled, metals (Fermi liquids) can be understood using rather simple classical models~\cite{nozieres1999theory} and it can be summarized simply as the presence of a sound mode dressed by electromagnetic interactions. It is important to clarify that the sound mode we are discussing is the electronic sound, usually denoted as zero sound. Strictly speaking, this term refers to the sound mode related to the zero temperature shape deformations of the Fermi surface. This hydrodynamic mode can be obtained using Fermi liquid theory~\cite{PhysRev.159.161} and it is just a manifestation of the elastic property of the electronic liquid~\cite{PhysRevB.60.7966}. Here, we use the term ``zero sound`` in a broader sense, to simply define the collective sound mode of the quantum critical soup both at zero and finite temperature. Despite the similar nature, it is not the normal sound associated with the vibrational modes of the ionic lattice, \ie the standard phonons. We will discuss this point further in section~\ref{sec:breaking2}.
\\
The underlying picture is less clear when more ``exotic'' phases are considered. In particular, recent experiments~\cite{KNUPFER1994121,PhysRevLett.59.2219,PhysRevB.43.3764,PhysRevB.39.12379,1991PhRvB..44.7155N} examined the dynamics of the collective plasmon modes in strongly correlated and quantum critical materials. The results are quite different from the weakly coupled paradigm, in the sense that the plasmon modes display an anomalously strong damping. Not only that, but there are recent observations~\cite{Mitrano5392,husain2019crossover} which indicate that when increasing the momentum, plasmons stop existing as well-defined quasi-particles and get smoothed out in the collective and incoherent ``quantum soup'' losing a characteristic momentum scale. The strong damping of plasmons, even at zero momentum, is expected for strongly coupled quantum critical systems and was first observed in a holographic model in~\cite{Aronsson:2017dgf}.
This was later elaborated on in~\cite{Krikun:2018agd}, which promoted the idea that the plasmons are no longer kinematically protected, since the electron-hole (Lindhard) continuum is now substituted by a continuum of modes typical of quantum critical systems. As a consequence, the plasmons can easily decay into such incoherent set of states and therefore display an anomalous and new damping mechanism (even at zero momentum).

The previous discussion leads us to an important point of this paper which is the identification and the analysis of the dielectric relaxation and plasmons damping mechanisms in a strongly coupled medium, as \eg in a quantum critical phase of matter. Working with linear response, and assuming zero momentum for simplicity, the complex dielectric function $\epsilon(\omega)$ controls the relation between the electric field $E$ and the polarization $P$,
\begin{equation}
    P(\omega)\,=\,\epsilon_0\,\left[\epsilon(\omega)\,-\,1\right] E(\omega)\,,
\end{equation}
Writing this relation in position space,
\begin{equation}
    P(t)\,=\,\epsilon_0\, \int \left[\epsilon(t-s)\,-\,\delta(t-s)\right] E(s) \, ds\,,
\end{equation}
it becomes manifest that a non-trivial dielectric function implies a delay, \ie relaxation mechanism, between the electric field and the polarization in the material.

The simplest phenomenological model for the dielectric constant is the well-known Debye model. It relies on a single relaxation time approximation for the polarization:
\begin{equation}
    \frac{d P(t)}{dt}\,=\,-\,\frac{1}{\tau_D}\,P(t)\,,
\end{equation}
where the timescale $\tau_D$ is denoted as the Debye relaxation time. Within this model, the dielectric function can be written simply as:
\begin{equation}
    \epsilon(\omega)\,=\,\epsilon_{\infty}\,+\,\frac{\Delta \epsilon}{1\,+\,i\,\omega\,\tau_D}\,,
\end{equation}
where $\epsilon_{\infty}$ is its value at infinite frequency.
The Debye model is justified by a basic molecular framework~\cite{doi:10.1002/9780470142806.ch5}, but it still clearly represents an approximation and will be insufficient when more involved processes are involved~\cite{Hill_1985}. It thus stands to reason that particularly in presence of strong coupling and collective dynamics the Debye model has to be extended, and one of the main goals of this paper is to provide further insight into how this extension has to be established.
Moreover, along the lines of~\cite{Aronsson:2017dgf, Krikun:2018agd}, it is important to mention that in the quantum critical systems considered, \eg strange metals, the electric conductivity displays very peculiar features. In particular, it differs from the standard wisdom (Drude model) and it reads schematically:
\begin{equation}
    \sigma(\omega)\,=\,\sigma_0\,+\,\sigma_{Drude}(\omega)~.
\end{equation}
The first term is an important new contribution, denoted as the \textit{incoherent conductivity}~\cite{Davison:2015taa}, which is not present in Galilean invariant systems and is definitely not considered in the Drude model for free electrons. It is completely insensitive to momentum dissipation and it can be thought of as a quantum critical contribution. From a more microscopic perspective, it can be generated from the production of particle-hole excitations carrying charge but no net total momentum. As a matter of fact, given the relation between the electric conductivity and the dielectric constant~\eqref{rel}, it is not suprising that this new term can induce new relaxation phenomena and therefore new contributions to the damping of the plasmons.

In summary, in this manuscript, we aim to investigate in details the relaxation mechanism and the damping of the quantum critical plasmons using simple holographic bottom-up models. We will first discuss the effects of polarization and the similarities/differences with the simple Debye relaxation model. Then, we analyze carefully the dispersion relation of the quantum critical plasmons and in particular their possible overdamped nature. We will discover a peculiar transition between a standard plasmon dispersion relation $\omega^2\,=\,\omega_p^2\,+\,c^2 k^2$ and a propagating sound wave $\omega=c k$ which in the intermediate regime displays a $k$-gap dispersion relation~\cite{Baggioli:2019jcm,Grozdanov:2018fic}\footnote{This behavior is quite general for the longitudinal fluctuations, but it has recently discussed also for the transverse sector in specific models ~\cite{Baggioli:2018vfc,Baggioli:2018nnp,Grozdanov:2018ewh}. One of them is, for example, the linear axions model we consider in section \ref{sec:breaking}.}. The $k$-gap simply refers to a dispersion relation of the type:
\begin{equation}
    Re[\omega]\,\sim\,c\,\sqrt{k^2\,-\,k_g^2}\label{eq:kgap}
\end{equation}
where a propagating mode appears beyond a certain cutoff value $k=k_g$ (see ~\cite{Baggioli:2019jcm}).\\

Finally, we investigate the role of broken translational invariance on the plasmon dynamics. This last part is entirely new and it is motivated by several recent experimental observations~\cite{principi2013impact,viola2018graphene,doi:10.1063/1.348957,doi:10.1021/jp504467j} and by the need of making the holographic models more realistic. More precisely, we analyze two different situations:
\begin{itemize}
    \item The introduction of momentum dissipation.
    \item The introduction of phononic degrees of freedom via the spontaneous symmetry breaking of translations.
\end{itemize}
In order to realize these two patterns we will use a simple holographic massive gravity model introduced in~\cite{Baggioli:2014roa,Alberte:2015isw} and studied in several directions in~\eg~\cite{Baggioli:2015zoa,Baggioli:2015gsa,Baggioli:2015dwa,Baggioli:2016oqk,Alberte:2016xja}.

\begin{figure}[t!]
    \centering
    \tcbox[colframe=green!30!black,
           colback=white!30]{ \includegraphics[width=0.9\linewidth]{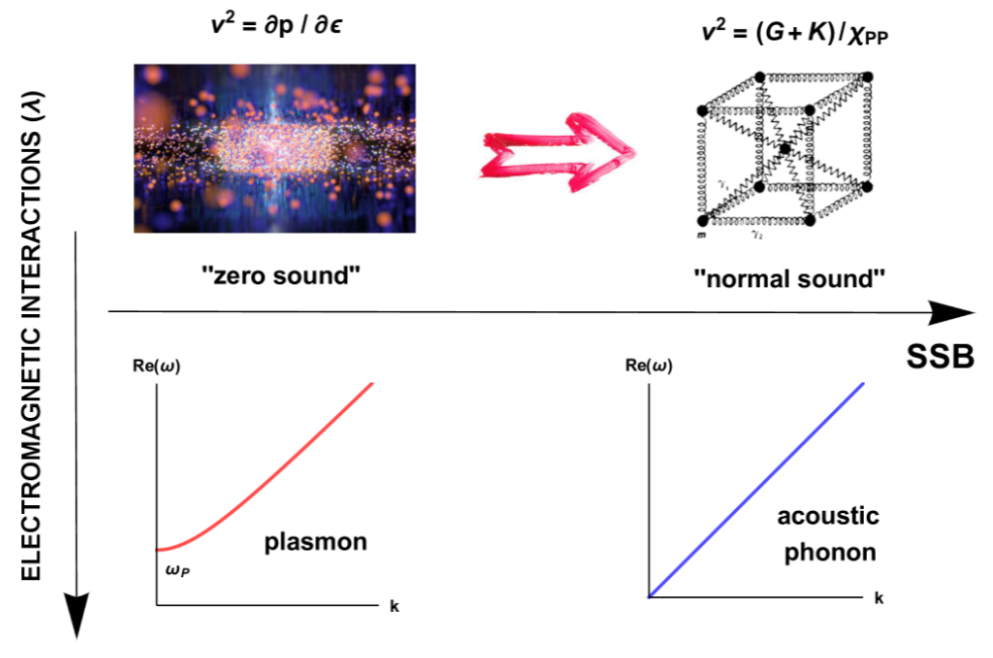}}
    \caption{A simplified pictorial representation of the ``zero sound'' to ``normal sound'' crossover when increasing the strength of the SSB. The ``zero sound'', appearing in quantum electron systems like Fermi liquids, gets gapped by the electromagnetic interactions producing the plasmon mode. On the contrary in the ``normal sound'' regime, typical of an ordered ionic crystal, the electromagnetic interactions do not gap the sound mode. In our system, the unique longitudinal sound mode interpolates between these two regimes by increasing the strength of the SSB.}
    \label{figsketch}
\end{figure}
One very important point is to better understand the nature of the longitudinal sound in holography. Even in absence of spontaneous symmetry breaking of translations, the longitudinal sector of the fluctuations display a sound mode which has been identified as the strongly coupled version of the zero sound of Fermi liquids~\cite{Karch:2008fa,Kulaxizi:2008kv,Edalati:2010pn,Karch:2007pd,HoyosBadajoz:2010kd}. Once SSB of translations is introduced in the model, the nature of that sound mode changes. In particular, the SSB provides a normal sound contribution which is going to be added to the original ``zero sound'' mode. This can be directly seen in the value of the speed of such a mode. It gets an additional contribution which is proportional to the elastic moduli which becomes more and more dominant at strong SSB. In other words, dialing the strength of the SSB the sound mode crosses over from a ``zero sound'' nature to a normal sound one (see figure~\ref{figsketch}). By studying the effects of the electromagnetic interactions on this sound mode we can indeed confirm this picture. While at small SSB, the sound mode gets gapped by the polarization effects, at large SSB it does not. This is indeed the indication that the sound mode becomes the normal sound of the elastic phonons, which, as we know, is not affected by the electromagnetic interactions.
Our findings provide new insights on the nature and the dynamics of bulk plasmons and sound modes in the realm of strongly coupled and quantum critical metals.\\[0.0cm]

\textbf{Note added: } As this work was in the final stages, we became aware of~\cite{Romero-Bermudez:2019lzz} which also studies plasmon attenuation in a different holographic model with pseudo-spontaneous breaking of translations.

\section{The model}\label{secmodel}
We consider the elementary class of holographic models introduced in~\cite{Baggioli:2014roa,Alberte:2015isw} and described by the following action:
\begin{equation}\label{eq:action}
S\,=\, M_P^2\int d^4x \sqrt{-g}
\left[\frac{R}2+\frac{3}{\ell^2}- \,\frac{1}{4}\,m^2 \,V(X)\,-\,\frac{1}{4}\,F^2\right]~,
\end{equation}
where $M_P$ is the Planck energy scale and $l$ the AdS radius\footnote{Both these quantities are fixed to unit in the computations.}. Additionally, we define $X \equiv  \, g^{\mu\nu} \,\partial_\mu \phi^I \partial_\nu \phi^I$ and the field strength of the $U(1)$ bulk field, $F=dA$. We study 4D asymptotically AdS black hole geometries whose metric reads:
\begin{equation}
\label{backg}
ds^2=\frac{1}{z^2} \,\left[\,-f(z)\,dt^2\,+\,\frac{dz^2}{f(z)}\,+\,dx^2\,+\,dy^2\,\right]~,
\end{equation}
where $z\in [0,z_h]$ is the radial holographic direction spanning from the boundary to the horizon, defined through $f(z_h)=0$. The $\phi^I$ are the St\"uckelberg scalars which admit a radially constant profile $\phi^I=\alpha\, x^I$ with $I=x,y$ (see~\cite{Alberte:2015isw} for more details) and $A_t=h(z)=Q(1-z/z_h)$ is the background solution for the $U(1)$ gauge field encoding the chemical potential $\mu$ and the charge density $\rho$ as
\begin{equation}
    \mu			\;=\;	\frac{Q}{\sqrt{\lambda}\, z_h}	\, , \quad
\rho	\;=\;	\frac{\sqrt{\lambda}\, Q}{z_h^2}	\, ,
\end{equation}
where the parameter $\lambda$ plays the role of the electromagnetic coupling $e^2$, determining the strength of the long range Coulomb interactions~\cite{Aronsson:2017dgf}.
The blackening factor takes the simple form
\begin{equation}\label{backf}
f(z)= z^3 \int_z^{z_h} dv\;\left[ \frac{3}{v^4} -\frac{m^2}{4\,v^4}\, 
V(v^2)\,-\,\frac{Q^2}{2 z_h^4} \right] \, ,
\end{equation}
and it vanishes at the horizon location $z=z_h$.
The corresponding temperature of the dual QFT reads
\begin{equation}
T=-\frac{f'(z_h)}{4\pi}=\frac{12 -  m^2\, V\left(z_h^2 \right)\,-\,2\,Q^2 }{16\, \pi \,z_h}~,\label{eq:temperature}
\end{equation}
while the entropy density is simply $s=2\pi/z_h^2$. Due to the symmetries of the system only dimensionless quantities are meaningful. For simplicity, without loss of generality, we fix $z_h=1$ and always refer to dimensionless quantities e.g. $\omega/T$, $\mu/T$ along the paper.

Depending on the choice of the potential $V(X)$, the model defined in~\eqref{eq:action} realizes different translational symmetry breaking patterns~\cite{Alberte:2017cch,Alberte:2017oqx,Ammon:2019wci}.
In this manuscript we consider the following cases:
\begin{enumerate}
\item $m^2=0, \,\mu=0$ : Section~\ref{sec:schwarzschild}.\\
This corresponds to a Schwarzschild black hole geometry. In this case the dynamics of the gauge and gravitational sectors decouple.
\item $m^2=0,\,\mu \neq 0$ : Section~\ref{secRN}.\\  This scenario reduces to the Reissner--Nordstr\"om (RN) background solution.
\item $m^2\neq 0,\,\mu \neq 0$ and $V(X)=X$ : Section~\ref{sec:breaking}.\\
This corresponds to the well-known \textit{linear axions model}~\cite{Andrade:2013gsa}.
\item $m^2\neq 0,\,\mu \neq0$ and $V(X)=X^3$ : Section~\ref{sec:breaking2}.\\
For this choice of potential the model exhibits spontaneous symmetry breaking (SSB) of translational invariance and a finite elastic response~\cite{Alberte:2017oqx}.
\end{enumerate}
We consider the longitudinal sector of the fluctuations (see appendix~\ref{app1} for more details). In order to take into account the effects of Coulomb interactions and polarization, we follow closely the method introduced in~\cite{Aronsson:2017dgf, Aronsson:2018yhi}. More precisely, we observe that the fluctuations of the gauge field $A_\mu$ behave close to the UV AdS boundary ($z=0$) as:
\begin{equation}
    \delta A_\mu\,=\,\delta A_\mu^{(0)}\,+\,\delta A_\mu^{(1)}\,z\,+\,\mathcal{O}(z^2)~.
\end{equation}
We can now convert the `plasmon condition' (\ref{rel0}) into a modified boundary condition for the gauge field at $z=0$, by using the relation to the conductivity in (\ref{rel}). This is straightforward to express in terms of the bulk fields using the holographic dictionary, c.f.~\cite{Aronsson:2017dgf, Aronsson:2018yhi}.
This results in the following boundary condition
\begin{equation} 
   \left( \omega^2\,\delta A_x^{(0)}\,+\,\lambda\,\delta A_x^{(1)}\,\right)\,=\,0\,,\label{eq:bcs}
\end{equation}
assuming, without loss of generality, that momentum is in $x$-direction, and Dirichlet boundary conditions for the other components. The original idea can be found in~\cite{Aronsson:2017dgf} and it has been used extensively in the successive studies of plasmons in holography~\cite{Aronsson:2018yhi,Mauri:2018pzq,Gran:2018vdn,Gran:2018jnt,Krikun:2018agd,Romero-Bermudez:2019lzz}. 
In the rest of the manuscript we will use the boundary condition~\eqref{eq:bcs} for the gauge field fluctuations and the standard Dirichlet boundary conditions for the other bulk field fluctuations.

We are interested in the density response of the holographic model which is encoded in the density-density correlator:
\begin{equation}\label{zero}
    \chi^0(\omega,k)\,\equiv\,\langle \rho(\omega,k)\,\rho(\omega,-k)\,\rangle\,,
\end{equation}
where $\rho$ is identified with the time component of the current operator $J^\mu$, dual to the bulk gauge field $A_\mu$. 

Introducing the deformed boundary conditions~\eqref{eq:bcs} corresponds to gauging the U(1) symmetry of the boundary field theory~\cite{Witten:2003ya}. From a technical point of view, this procedure is equivalent to a double trace deformation of the boundary field theory~\cite{Witten:2001ua, Mueck:2002gm, Aronsson:2017dgf, Mauri:2018pzq, Krikun:2018agd}.

Taking an alternative view, we obtain the physical response function $ \chi(\omega,k)$ by dressing the density-density correlator via the electromagnetic interactions induced by the, now dynamical, photon. The physical response function can be defined as:
\begin{equation}
    \chi(\omega,k)\,=\,\frac{\chi^0(\omega,k)}{1-\,V_k\,\chi^0(\omega,k)}\,,\label{final}
\end{equation}
where $V_k \equiv \lambda/k^2$, with $\lambda$ the electromagnetic coupling appearing in the boundary condition~\eqref{eq:bcs}.
This last result reproduces correctly the outcomes of the RPA approximation~\cite{nozieres1999theory, Zaanen:2015oix}. In this language, the plasmons are simply the poles of the physical response function, or dressed density-density correlator, and they can be substantially different from the poles of the screened response function~\eqref{zero}, which by definition are the quasinormal modes of the system, see e.g.~\cite{Kovtun:2005ev} for more details.

In summary, the main object of our discussion will be the physical response encoded in eq.~\eqref{final} and the collective excitations appearing therein.
\section{The effects of the electromagnetic interactions}\label{sec:schwarzschild}
We start by considering the model described in eq.~\eqref{eq:action} in absence of a background charge density, $\rho=0$, and in absence of the translations breaking scalar sector, $m^2=0$. This choice corresponds to a vanishing background profile for the $U(1)$ bulk gauge field $A_\mu$. We will study the effects of the electromagnetic interactions through the modified boundary conditions~\eqref{eq:bcs} with $\lambda \neq 0$.

\begin{figure}[t!]
\centering
\includegraphics[width=0.49\linewidth]{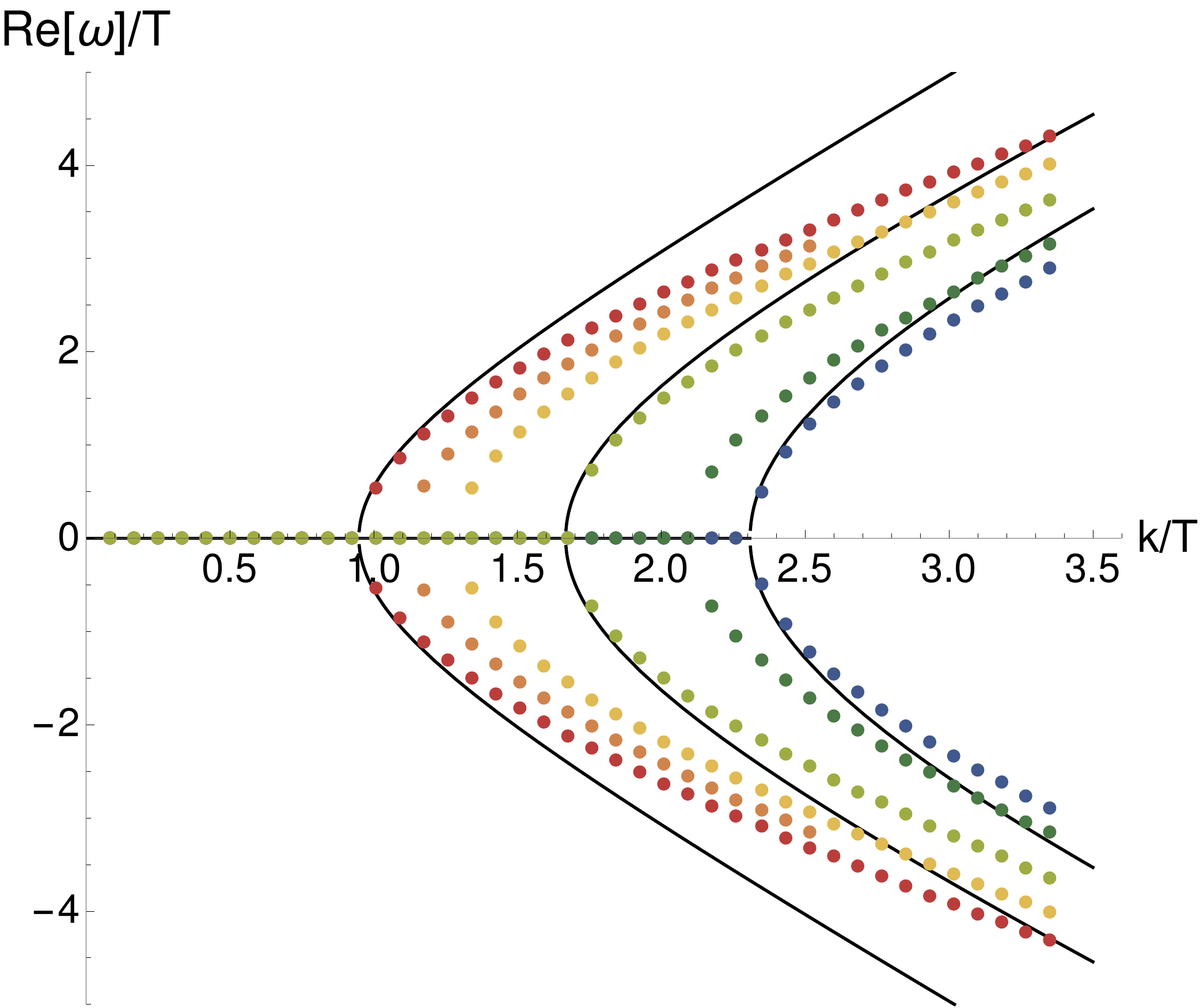}
\includegraphics[width=0.49\linewidth]{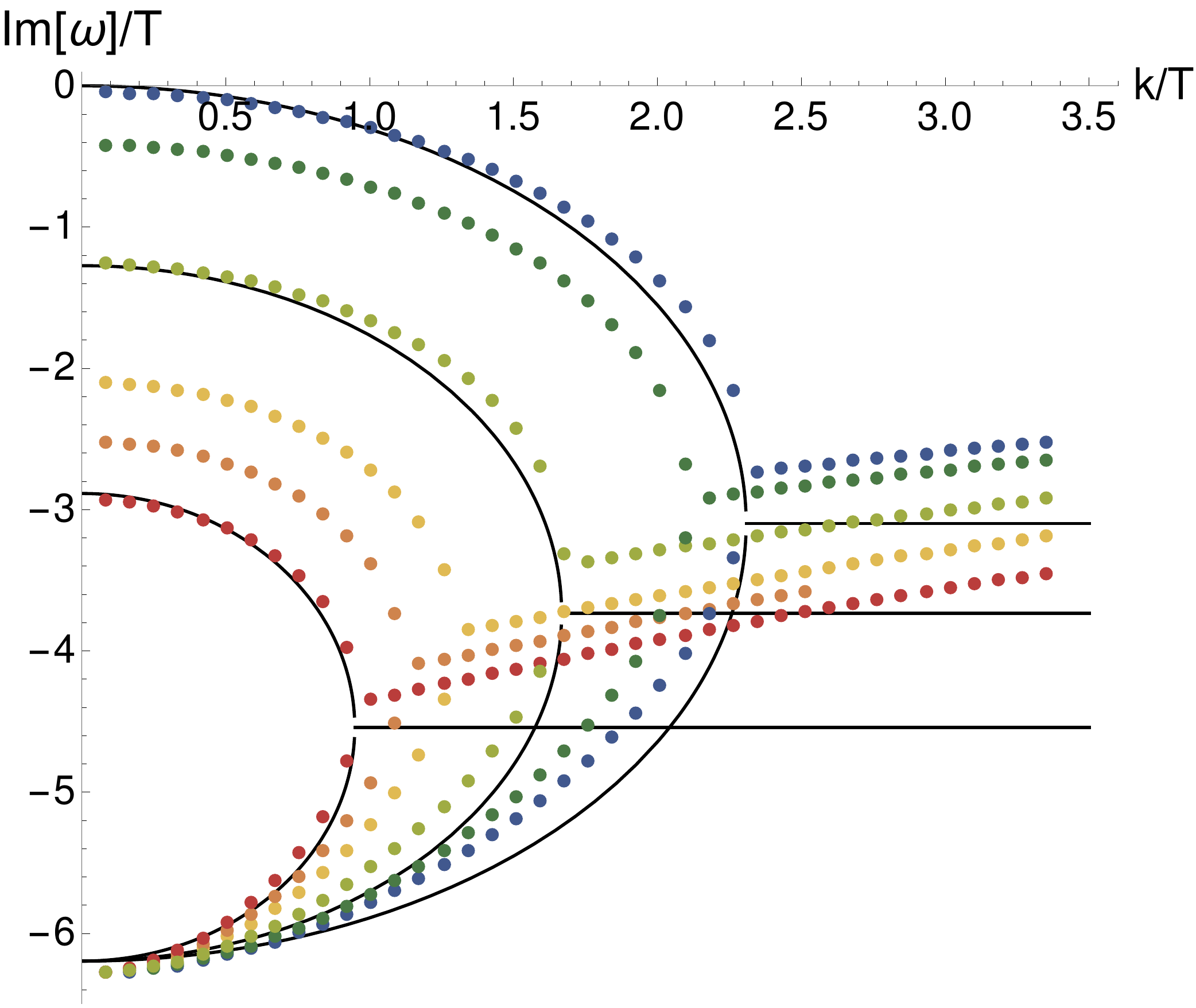}
\caption{Real and imaginary parts of the lowest modes with $\lambda{[0.01-0.7]}$ (blue-red). The blacks lines show some fits using formula~\eqref{tre}. The dispersion relations show the appearance of a propagating mode above a certain momentum $k\equiv k_g$ typical of the $k$-gap phenomenon.}
\label{fig1}
\end{figure}

Before proceeding, let us remind the reader about the result for zero charge density and standard Dirichlet boundary conditions, \ie $\lambda=0$. Importantly, in this limit the gauge and the gravitational sectors decouple and their dynamics are completely independent. In the gravitational sector, the only hydrodynamic pole in the longitudinal spectrum is a sound mode:
\begin{equation}
    \omega\,=\,v_s\,k\,-\,i\,D\,k^2\,+\,\dots
\end{equation}
The sound speed can be computed from the standard thermodynamic relation $v_s^2=dp/d\varepsilon$ with $p$ the pressure and $\varepsilon$ the energy density of the system. As a consequence of the conformal invariance of the system, the speed is constrained to take the simple value $v_s^2=1/(d-1)$. At the same time, the sound attenuation reads $D=\eta/(2s T) = 1/(8 \pi T)$~\cite{Policastro:2002se,Policastro:2002tn,Kovtun:2012rj}.
On the other hand, the decoupled gauge sector displays a single hydrodynamic diffusive mode which arises from the conservation of the electric current. More specifically, the charge diffusion mode reads:
\begin{equation}
    \omega\,=\,-\,i\,D_Q\,k^2\,,\quad D_Q\,=\,\frac{\sigma}{\chi_{\rho\rho}}~,\label{uno}
\end{equation}
where $\sigma$ is the electric conductivity and $\chi_{\rho\rho}\equiv \partial \rho/\partial \mu$ the charge susceptibility. For the background considered here we have $\sigma=1$ and $\chi_{\rho\rho}=1$. Therefore we find that:
\begin{equation}
    D_Q\,T\,=\,\frac{3}{4\,\pi}\,,
\end{equation}
which is in perfect agreement with our results in figure~\ref{fig1}. Notice how this diffusion constant does not depend on the electromagnetic coupling $\lambda$ at zero charge density.

If we extend the analysis to the first non-hydrodynamic mode\footnote{We use the standard definition of a hydrodynamic mode, namely a mode satisfying
$\lim_{k \rightarrow 0}\,\omega_{\text{mode}}(k)\,=\,0 ~$.
}, we find the presence of a second pole:
\begin{equation}
    \omega_1\,\sim\,-\,i\,\Gamma_1\,+\,i\,D_Q\,k^2\,+\,\dots\label{due}
\end{equation}
\\
The dynamics of the two modes appearing in eqs.~\eqref{uno} and~\eqref{due} can be described by the simple equation:

\begin{equation}
    \omega\,(\,\omega\,+\,i\,\Gamma_1\,)-\,D_Q\,\Gamma_1\,k^2\,=\,0~,
\end{equation}
which gives rise to the recently discussed $k$-gap phenomenon~\cite{Baggioli:2019jcm}. Schematically, the charge diffusion mode collapses at a specific wavelength with the first non-hydro mode and produces a propagating mode at low frequencies and large momenta\footnote{This behaviour appeared already in~\cite{Arias:2014msa}.}.\\

In order to compute the QNMs of the system we search for the zeroes of the determinant of the matrix valued bulk-to-boundary propagator \cite{Kaminski:2009dh}, see appendix \ref{app2} for a more explicit discussion. This corresponds to the blue dotted data shown in figure~\ref{fig1}. Once we introduce the electromagnetic interactions, by turning on $\lambda$,  the picture changes as shown in figure~\ref{fig1}.  The diffusive mode~\eqref{uno} acquires a finite damping at $k=0$. The two modes now satisfy a slightly different equation:
 \begin{equation}
        (\omega+i\,\Gamma_1)\,(\omega\,+\,i\,\Gamma_P)\,=\,c^2\,k^2\,,\label{tre}
    \end{equation}
where $\Gamma_P$ is the damping of the pseudo-diffusive mode and $c$ the asymptotic speed of the propagating mode. Let us remark that these are not the standard quasinormal modes~\cite{Kovtun:2005ev}. It is important to notice that in this sector there is no hydrodynamic mode.  The appearance of attenuation (at k=0) for the collective modes~\cite{Andrade:2017cnc} is associated to the non Fermi-liquid nature of the system (see~\cite{Krikun:2018agd} for a thorough discussion on this).  The damping  coefficient $\Gamma_P$ appearing in~\eqref{tre}, modelling the ``gapping'' of the diffusive mode, plays the role of a relaxation rate 
related to the effects of the electromagnetic interactions and the consequent polarization dynamics of the quantum critical sector. In some sense, this is a built-in feature of the holographic dictionary; as discussed in~\cite{Aronsson:2017dgf, Mauri:2018pzq}, the modified boundary conditions~\cite{Aronsson:2017dgf,Aronsson:2018yhi} correspond to a double trace deformation from the dual field theory point of view.  In presence of a double trace deformation, the corresponding operator (which is not anymore the U(1) conserved current) acquires an anomalous conformal dimension~\cite{Marolf:2006nd} and it is not conserved. Let us emphasize that the total charge density remains conserved. Formally, we expect that the presence of this relaxation time could be also understood in terms of the global symmetry formulation of dynamical electromagnetism introduced in~\cite{Grozdanov:2016tdf,Grozdanov:2018fic,Hofman:2017vwr}.\\
 \begin{figure}[t!]
\centering
\includegraphics[width=0.48\linewidth]{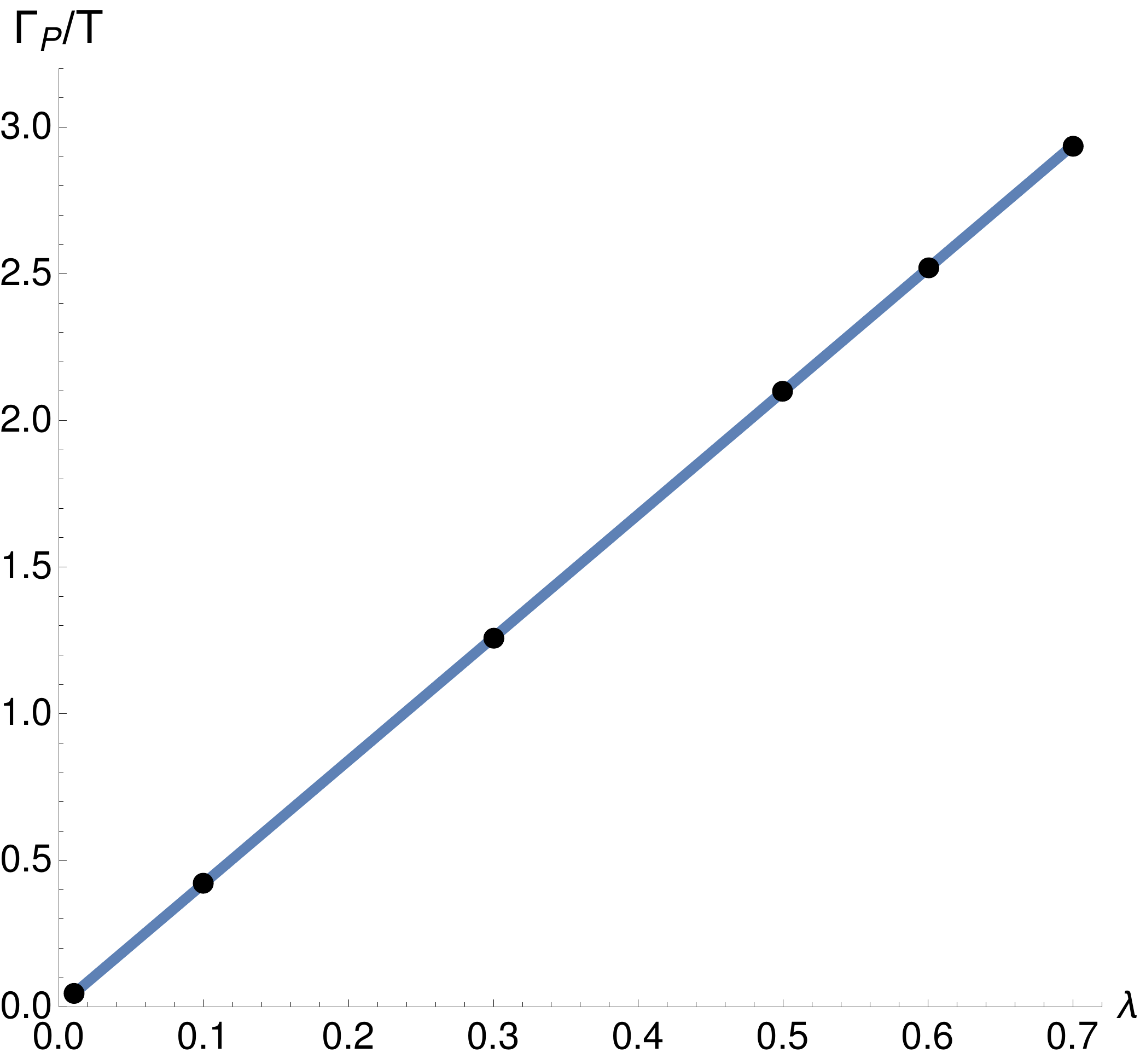}\quad
\includegraphics[width=0.48\linewidth]{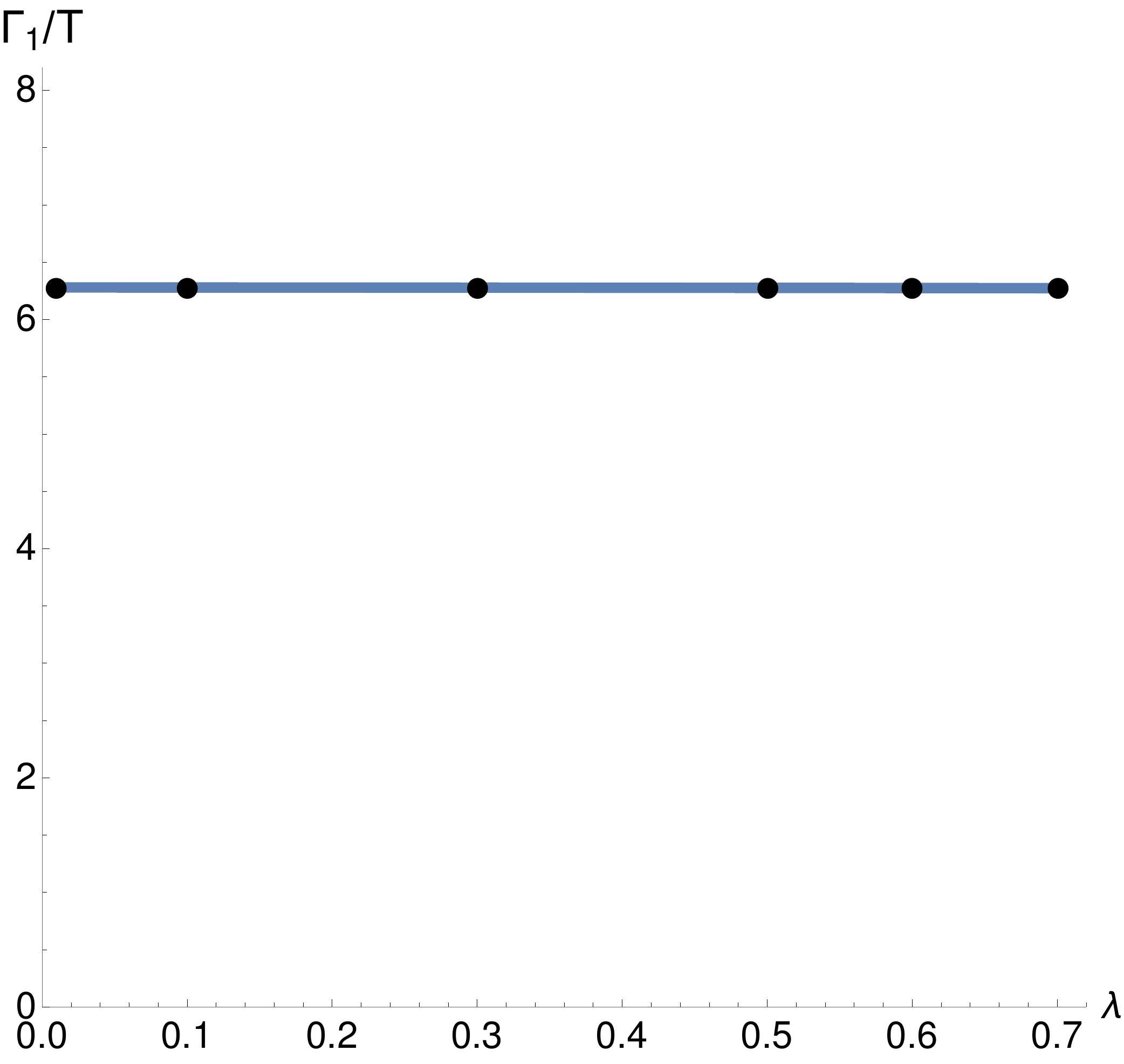}
\caption{The Debye and quantum critical damping coefficients entering in expression~\eqref{tre} in function of the electromagnetic coupling $\lambda$. The plot emphasize the very different nature of the two relaxation scales.}
\label{fig2}
\end{figure}\\
We are now interested in studying the dependence of the various parameters ($\Gamma_1,\Gamma_P,c$) as functions of the electromagnetic coupling $\lambda$ and the temperature $T$. The first observation, which is already evident in figure~\ref{fig1}, is that the damping parameter $\Gamma_1$ does not depend on the electromagnetic coupling. From our numerical data (see figure~\ref{fig2}) we observe that
\begin{equation}
    \Gamma_{1}\,=\,2\,\pi\,T~,
\end{equation}
which is a natural outcome since the temperature is the only dimensionful scale available. This frequency is just the first Matsubara frequency in a bosonic theory:
\begin{equation}
    \Gamma_n\,=\,2\,\pi\,T\,n~,
\end{equation}
where $n \in \mathbb{Z}$. We indeed find evidence for the higher Matsubara frequencies in our numerical data.
\\
 On the contrary, the other relaxation scale shown in figure~\ref{fig2}, is proportional to the electromagnetic coupling $\Gamma_P \sim \lambda$. This means that the associated relaxation mechanism is a direct effect of the long range Coulomb interactions parametrized by $\lambda$. Finally, let us spend a few words about the $k$-gap appearance and the momentum scale at which the two modes collide, \ie $k_g$. As is evident in figure~\ref{fig3} its value becomes smaller by increasing the electromagnetic coupling $\lambda$. As shown also in figure~\ref{fig1}, at higher values of $\lambda$ the collision appears at smaller values of the momentum. Nevertheless, this last feature does not imply that this collision becomes more and more hydrodynamic and relevant at late time. At large values of $\lambda$ both the modes are strongly damped and the collision appears at very large and negative imaginary values of the frequency.
\begin{figure}[t!]
\centering
\includegraphics[width=0.49\linewidth]{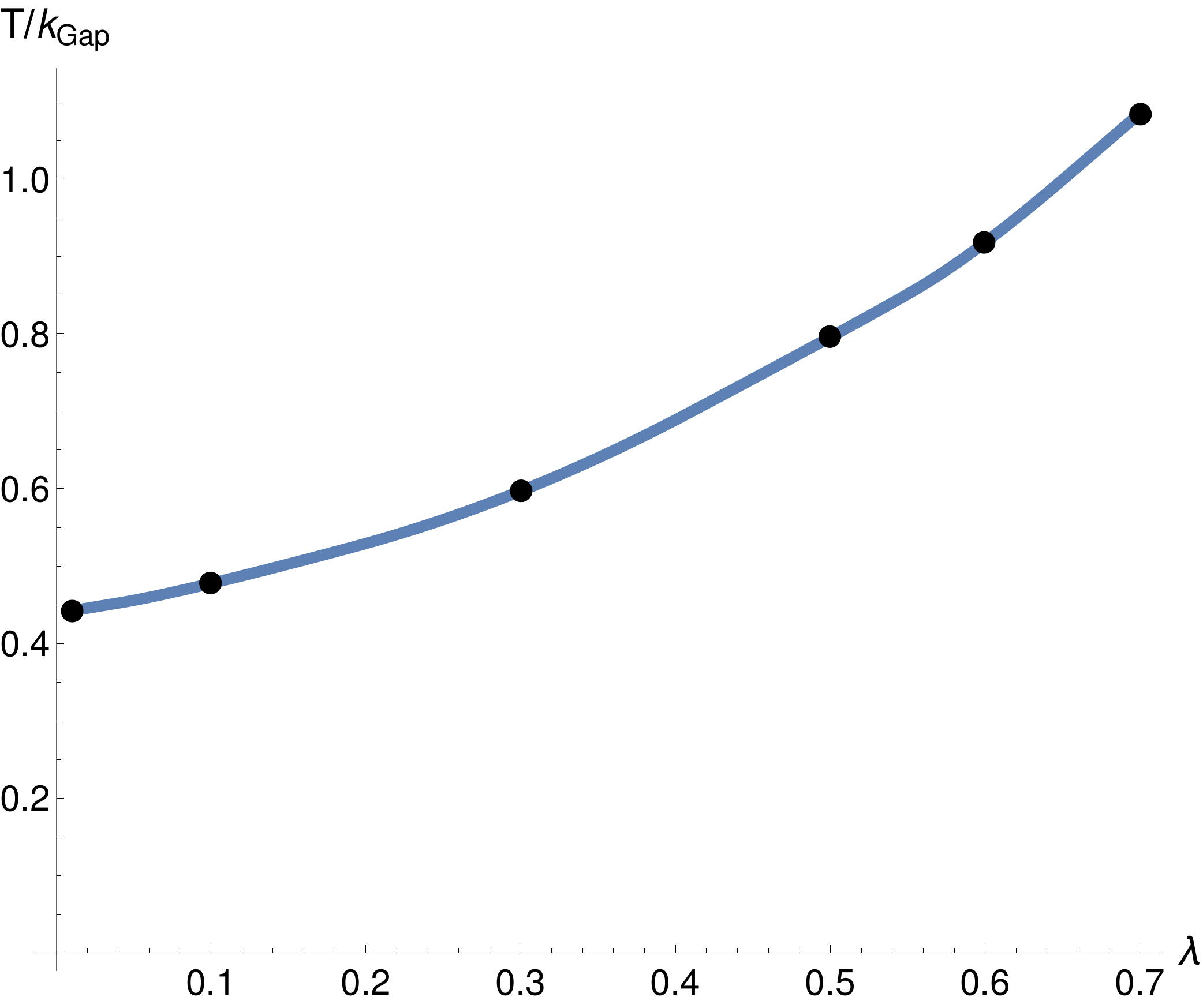}
\caption{The position of the $k$-gap as a function of the electromagnetic coupling $\lambda$. Increasing $\lambda$ the $k$-gap appears at smaller and smaller values of momentum.}
\label{fig3}
\end{figure}
\\
In summary, we have noticed that, even in the simple case of zero charge density, the holographic results go beyond the standard weakly coupled picture.
\section{Holographic plasmons at finite density}\label{secRN}
In this section we extend the computations of the previous part to the case of finite charge density, $\rho \neq 0$. At this stage, we still have $m^2=0$ and the system is still translational invariant.
In this scenario, the gravitational and gauge sectors are no longer decoupled. The $\mu=0$ limit is subtle. We emphasize that the analysis of the previous section~\ref{sec:schwarzschild} pertains to the gauge field sector and it does not take into account the decoupled gravitational dynamics.

In absence of Coulomb interaction, \ie $\lambda=0$, the longitudinal excitations of the standard Reissner--Nordstr\"om background have been intensively studied in the literature~\cite{Edalati:2009bi,Edalati:2010hk,Matsuo:2009yu}. In this setup, there are two hydrodynamic modes in the longitudinal spectrum: the sound mode and the charge diffusion mode. As shown in~\cite{Matsuo:2009yu}, their dispersion relations depend non-trivially on the dimensionless ratio $\mu/T$.

We now introduce the effect of long range electromagnetic interactions parametrized by the coefficient $\lambda$, \cf\eqref{eq:bcs}. The numerical results of this section were originally presented by some of the authors in~\cite{Gran:2018vdn}; here we interpret them in terms of a simple effective model.

We proceed by giving a phenomenological but analytic picture of what we expect, which is inspired by~\cite{Baggioli:2019jcm} and references therein.

Assuming a single relaxation time approximation, we can write the coherent part of the conductivity\footnote{It is important to notice that we are thinking about the electric conductivity in presence of Coulomb interactions. For a more detailed discussion see \eg\cite{Mauri:2018pzq}.} as:
\begin{equation}
\sigma(\omega,k)\,=\,\frac{n\,e^2\,\tau}{m^*}\,\frac{1}{1\,-\,i\,\omega\,\tau}\,+\,\dots \label{cond}
\end{equation}
where $n$ is the density of charge carriers, $m^*$ their effective mass, $e$ their charge and $\tau$ a characteristic relaxation timescale. The ellipsis indicate possible corrections which are subleading in the low frequency/momentum limit.
Let us emphasize that the relaxation time $\tau$ is a phenomenological parameter and its nature cannot be identified by any momentum dissipation mechanism. At this stage, momentum is a perfectly conserved quantity. This relaxation time is tied to the presence of electromagnetic interactions and possible polarization mechanisms and  it is present even at zero charge density. Technically it is simply a consequence of the double trace deformation encoded in the modified boundary conditions~\eqref{eq:bcs}. Moreover, at this stage, formula~\eqref{cond} is written in terms of quasi-particles, which are definitely absent in our system. We will amend this flaw later on.

Now, to make progress, we consider the fundamental equation for the longitudinal plasmons
\begin{equation}
    \epsilon_L(\omega,k)\,=\,0~,
\end{equation}
which was already introduced in the introduction (see~\cite{nozieres1999theory,Aronsson:2017dgf} for more details about its derivation).
At zero momentum, we can simply relate the dielectric function to the frequency dependent electric conductivity:
\begin{equation}
     \epsilon_L(\omega,0)\,=1\,+\,\frac{i\,4\,\pi}{\omega}\sigma(\omega)\,,\label{try}
\end{equation}
where momentum dependent corrections are not taken into account yet. The simplest form for the conductivity can be derived assuming a single relaxation time $\tau$, in analogy with the Drude model, and more specifically:
\begin{equation}
    \sigma(\omega)\,=\,\left(\frac{\omega_P^2\,\tau}{1\,-\,i\,\omega\,\tau}\right)\,,\quad \omega_P^2\,=\,\frac{4\,\pi\,e^2\,n}{m^*}~,
\end{equation}
where $\omega_P$ is the plasma frequency. At this point, the equations are written in terms of the plasma frequency and they do not rely on the existence of well defined quasi-particles. Now, we want to extend the expression~\eqref{try} to finite momentum. Given the invariance of the system under $x \rightarrow -x$, the first momentum correction has to be quadratic. More specifically, using dimensional analysis, we can write:
\begin{equation}
    \epsilon_L(\omega,k)\,=\,1\,+\,\frac{i}{\omega}\sigma(\omega)\,-\,\frac{c^2\,k^2}{\omega^2}\,+\,\mathcal{O}(k^4)~,\label{model2}
\end{equation}
where $c$ is a parameter with the dimensions of a velocity and the minus sign is chosen such that at $\sigma=0$ one recovers a standard sound wave $\omega= c k$.\\ All in all, at leading order in momentum, we obtain the final phenomenological expression\footnote{A similar formula was suggested in~\cite{Gran:2018vdn}.}:
\begin{equation}
\omega^2\,\left[1\,+\,\frac{\,i}{\omega}\,\left(\frac{\omega_P^2\,\tau}{1\,-i\,\omega\,\tau}\right)\,\right]\,=\,c^2\,k^2~,\label{model}
\end{equation}\\
which derives from two natural assumptions:
\begin{itemize}
    \item A single relaxation time approximation for the momentum independent electric conductivity $\sigma(\omega)$. This choice displays strong analogies with the Debye relaxation model for dielectric materials~\cite{doi:10.1002/9780470142806.ch5}.
    \item Considering only the first and quadratic momentum corrections to the longitudinal dielectric function $\epsilon_L(\omega,k)$.
\end{itemize}
 Given the order of the polynomial, we will have three different and interacting modes, which give rise to a quite complicated pattern. 
Let us analyze further the qualitative trend. At $\tau=0$, where dissipation is infinite, there is no screening and the dispersion relation is that of a simple propagating wave $\omega= c k$ with speed $c$. In the limit $\omega \tau \ll 1$ we obtain the simple equation:
\begin{equation}
\omega^2\,+\,\frac{i}{\omega}\,\omega_P^2\,\tau\,=\,c^2\,k^2\,,
\end{equation}
at leading order in $\omega \tau$. This is the well known expression producing the so-called $k$-gap dispersion relation~\eqref{eq:kgap}. In particular, the real part of the dispersion relation becomes non-zero only above a certain cutoff momentum, usually called $k$-gap $k\equiv k_g$, which can be easily derived as:
\begin{equation}
k_g\,=\,\frac{\omega_P^2\,\tau}{2\,c}\,.
\end{equation}
Increasing further the relaxation time $\tau$, the situation becomes more complicated and all the three modes interact creating a non-trivial interplay. Finally, in the limit of $\omega \tau \gg 1$, the dispersion relation becomes simply:
\begin{equation}
\omega^2\,=\,\omega_P^2\,+\,c^2\,k^2\,,
\end{equation}

which is the well-known gapped plasmon mode.

\begin{figure}[t!]
\centering
\includegraphics[width=0.40\linewidth]{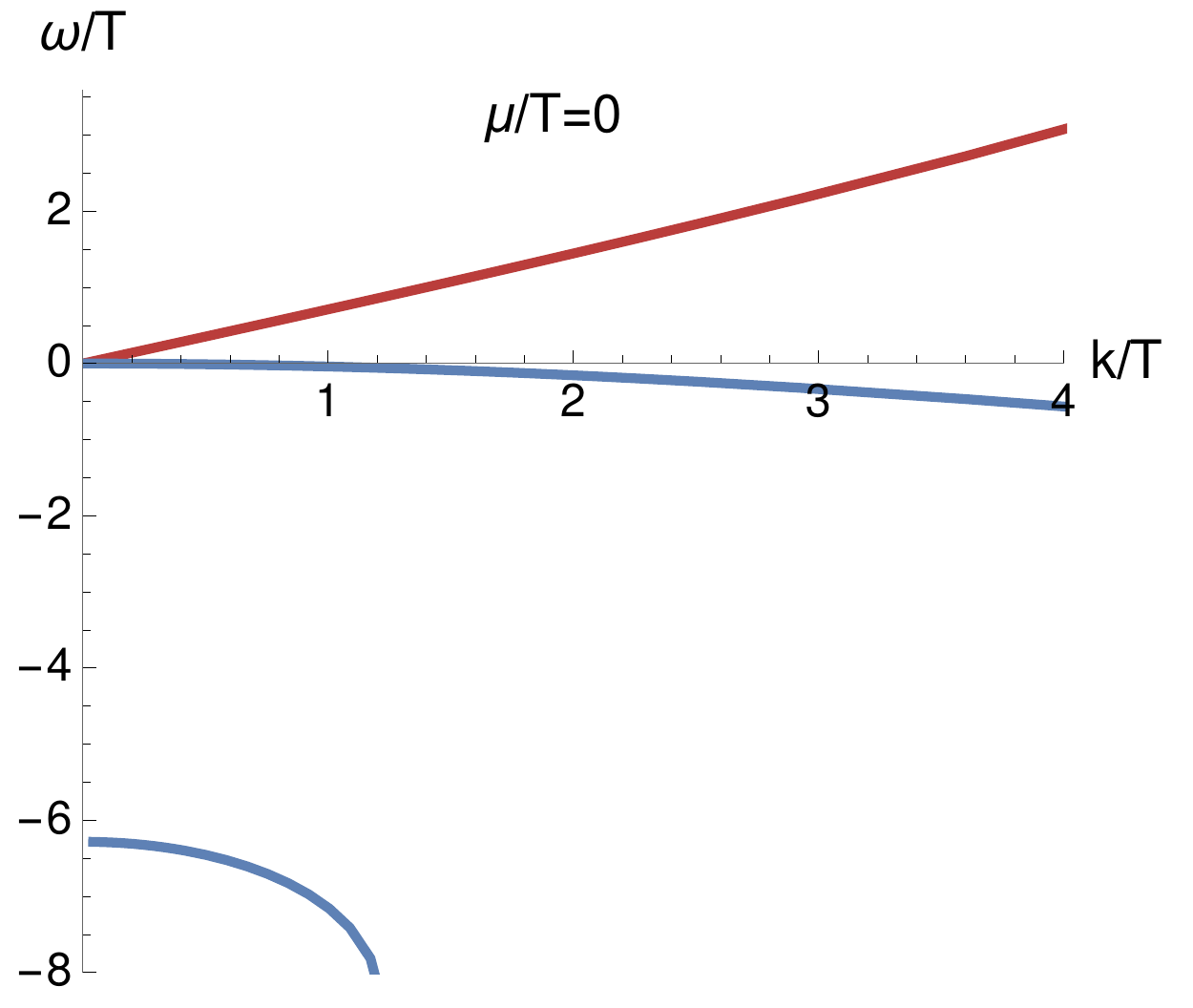}
\qquad 
\includegraphics[width=0.40\linewidth]{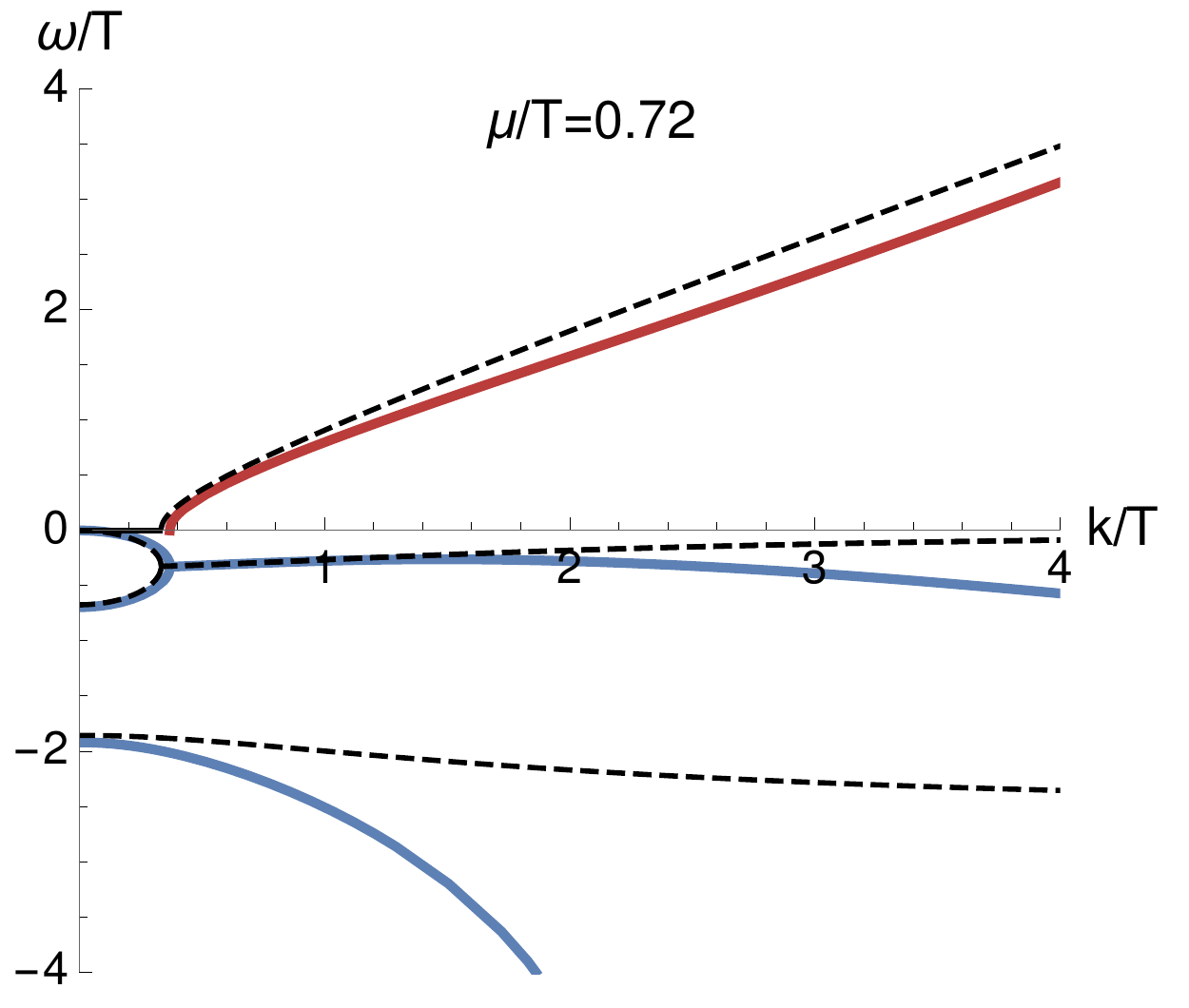}

\vspace{0.4cm}

\includegraphics[width=0.40\linewidth]{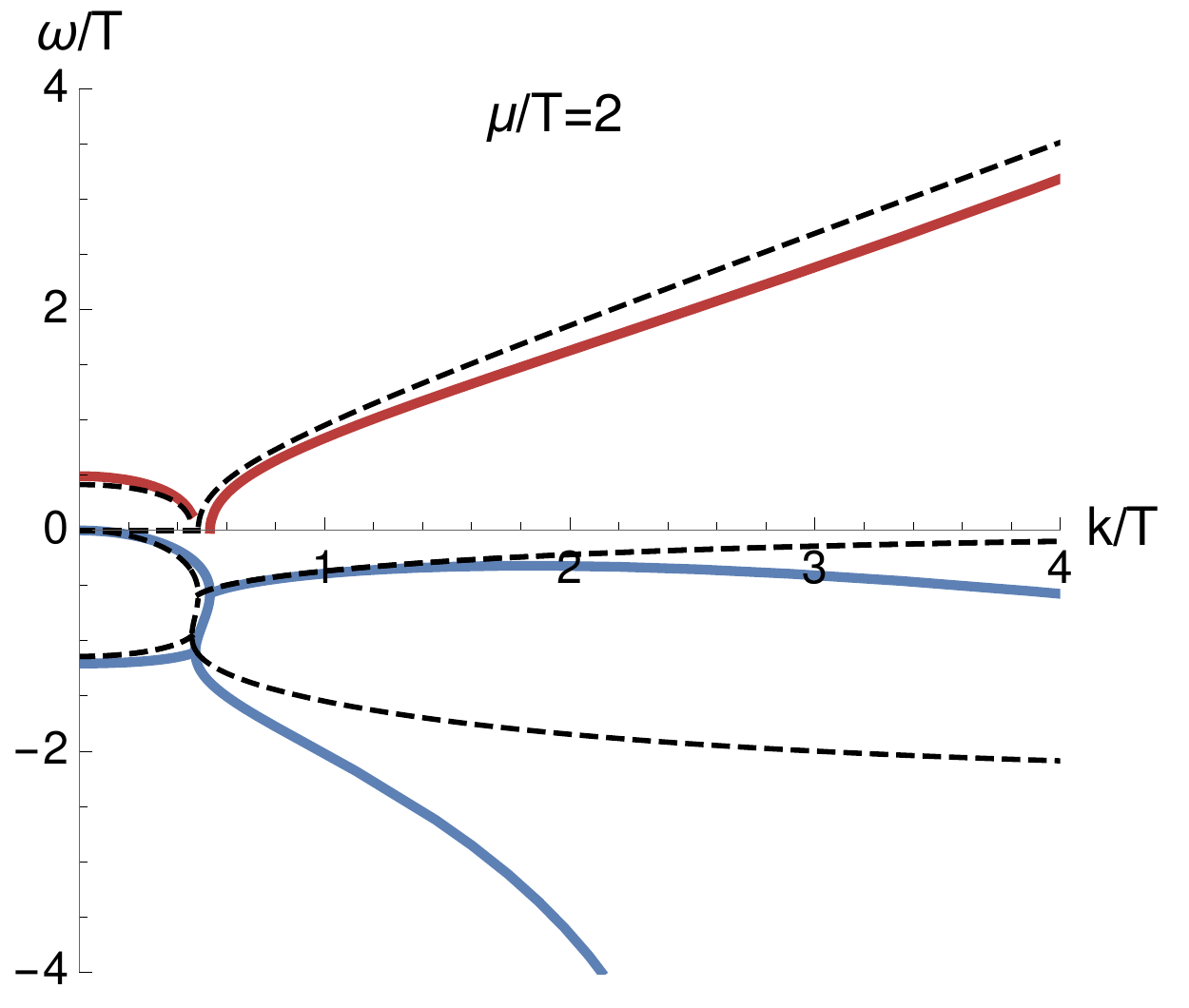}
\qquad
\includegraphics[width=0.40\linewidth]{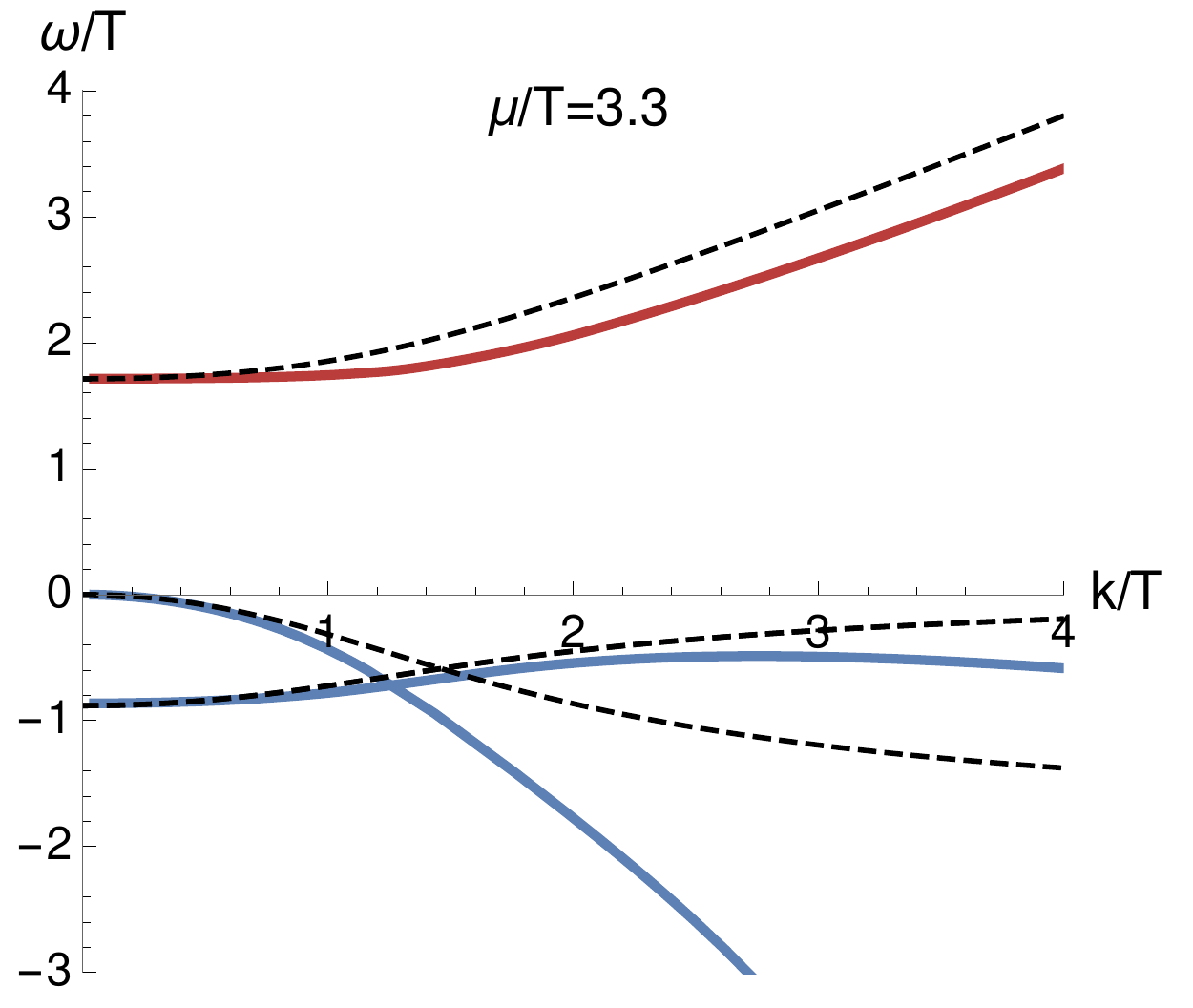}

\caption{ Real (red) and imaginary (blue) parts of the lowest, longitudinal, collective modes in the Reissner--Nordstr\"om geometry when increasing $\mu/T$ at fixed $\lambda=1$. Colored lines represent numerical data and dashed-black the fit to formula~\eqref{model}. The hydrodynamic formula fits well the data in the regime $\omega/T,k/T \ll 1$.}
\label{fig5}
\end{figure}

\begin{figure}[t!]
\centering
\includegraphics[width=0.40\linewidth]{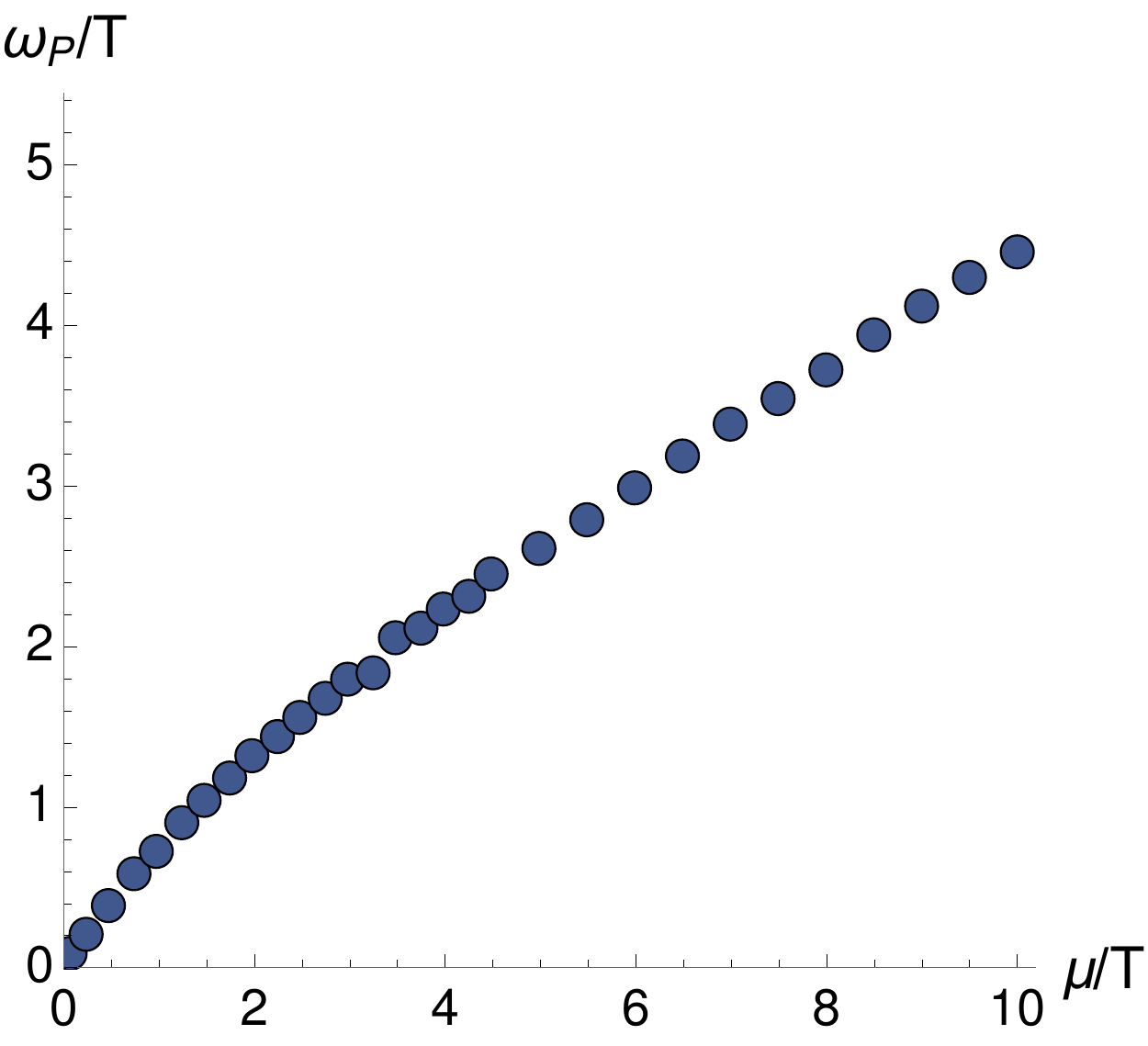}\quad
\includegraphics[width=0.40\linewidth]{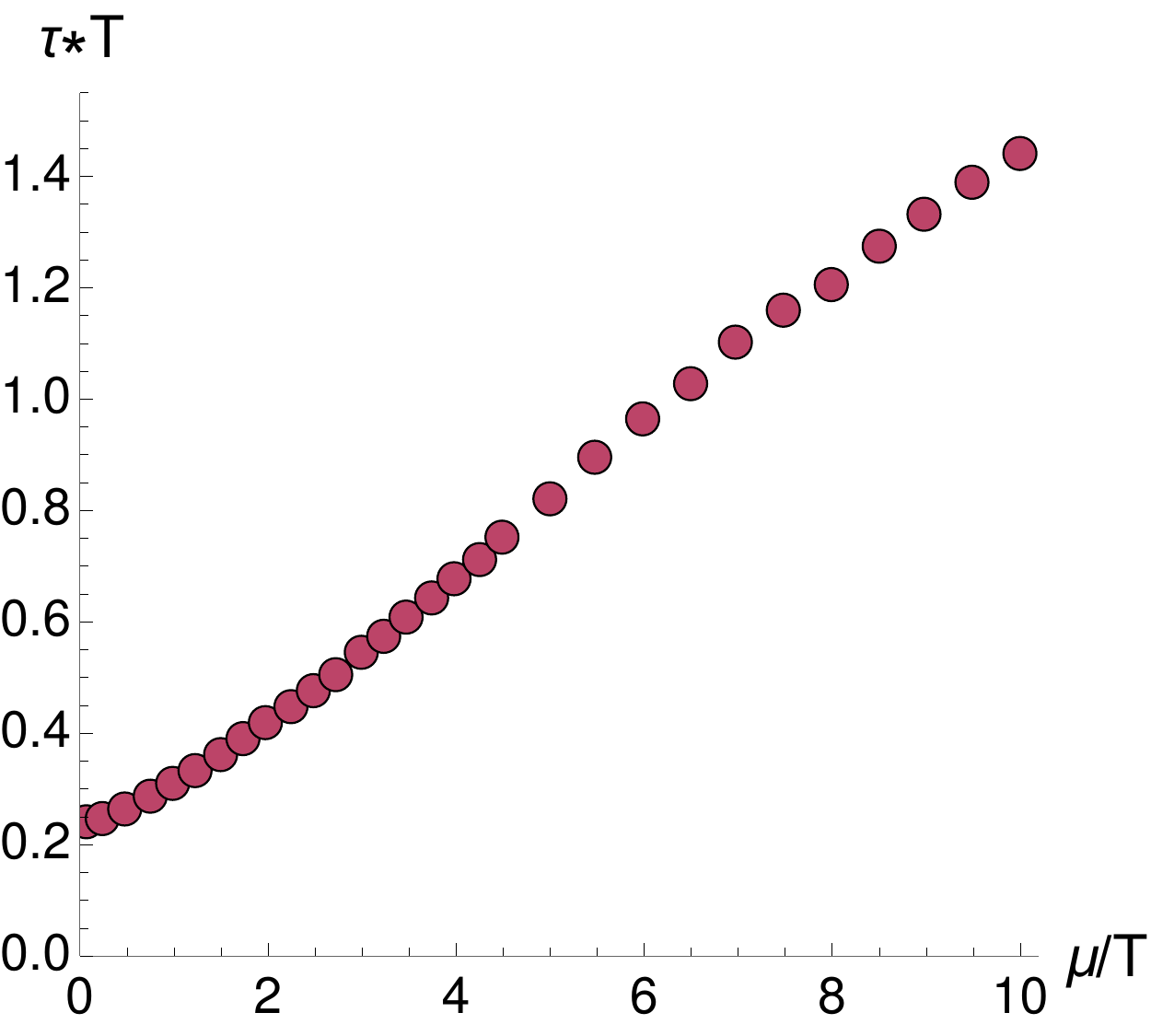}
\caption{Evolution of $\omega_P$ and $\tau$ with increasing $\mu/T$ for $\lambda=1$ obtained from the fits as shown in figure~\ref{fig5}.}
\label{fig:tauandwp}
\end{figure}

\begin{figure}[t!]
\centering
\includegraphics[width=0.40\linewidth]{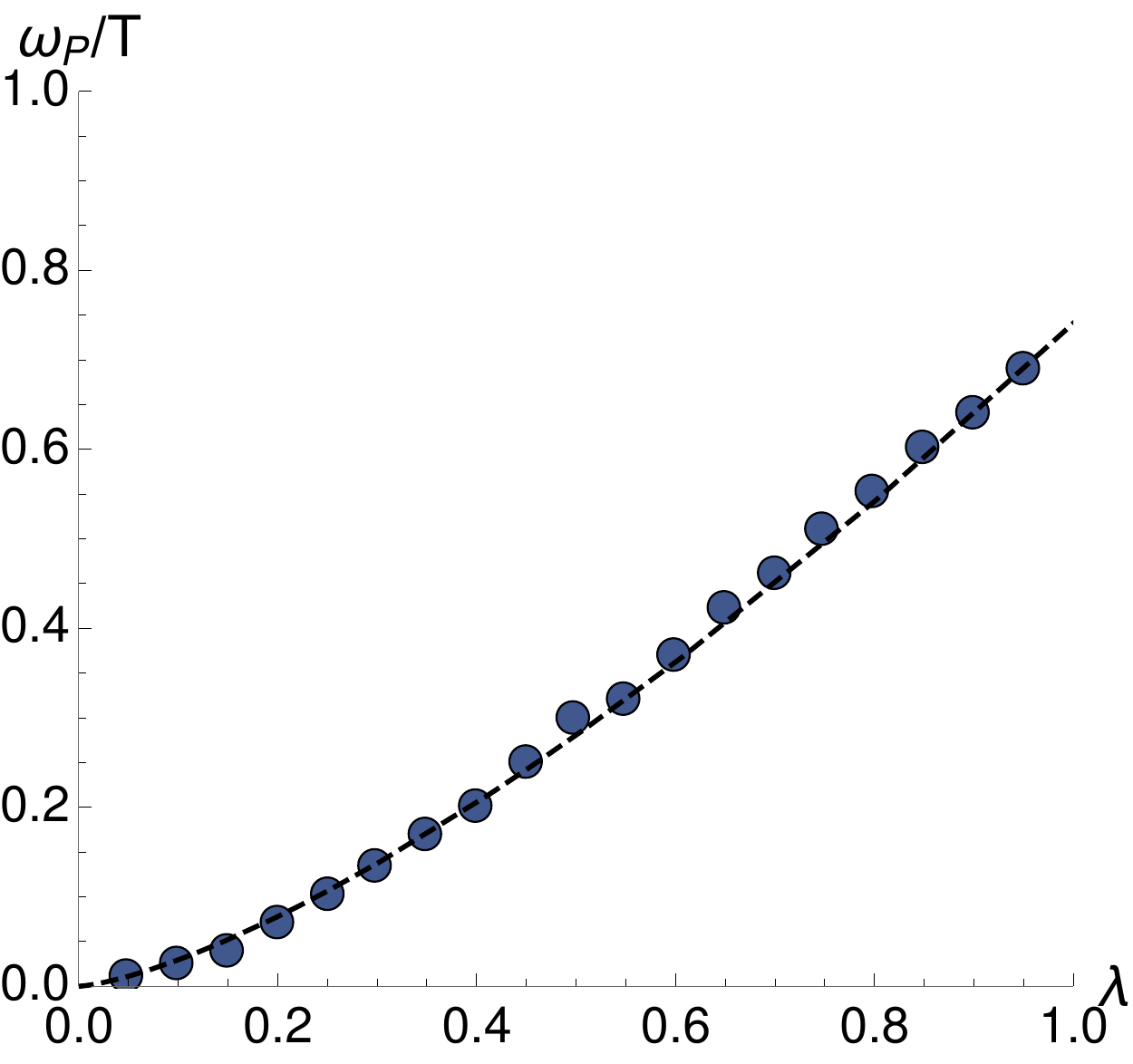}\quad
\includegraphics[width=0.40\linewidth]{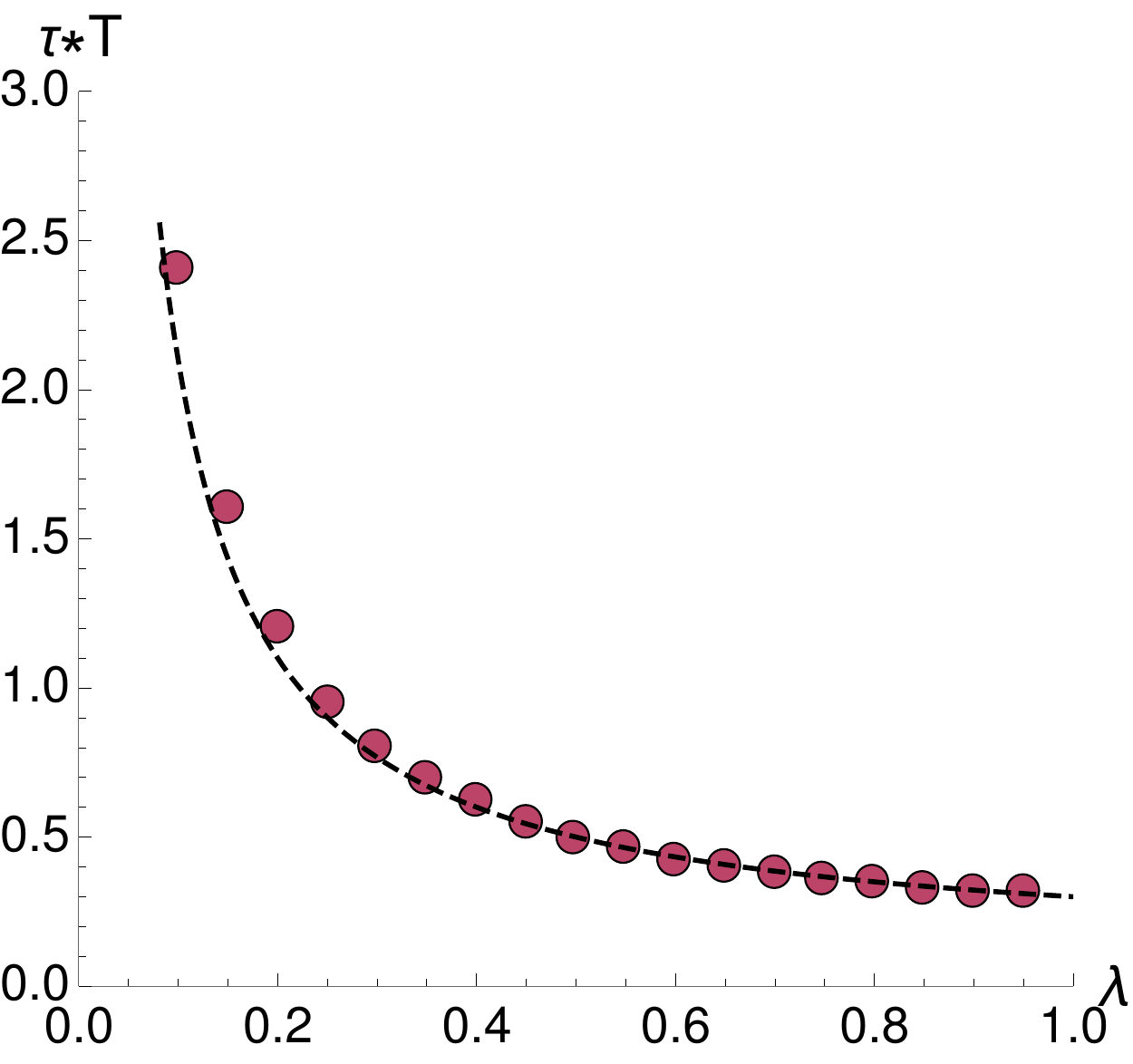}
\caption{Evolution of $\omega_P$ and $\tau$ as functions of $\lambda$ with fixed $\mu/T=1$ obtained from fitting the numerical data. The fit for the relaxation time $\tau$ (dashed line) is consistent with a linearly inverse scaling $\tau \sim \lambda^{-1}$.}
\label{fig:tauandwpoflambda}
\end{figure}

The main result of this section is shown in figure~\ref{fig5} where the numerical data are compared with the analytic predictions. The simple phenomenological expression~\eqref{model} captures successfully the dispersion relation of the three lowest modes and the complicated interplay between them at low frequency and momentum.  Moreover, as shown in figure~\ref{fig:tauandwp} we find that both the plasma frequency $\omega_p$ and the effective relaxation scale $\tau$ increase with the dimensionless parameter $\mu/T$. More precisely, the plasma frequency goes to zero with $\mu/T$ following a power-law, while the relaxation time approaches a finite value at $\mu/T=0$. The dependence of $\tau$ and $\omega_P$ on $\lambda$ is shown in  figure~\ref{fig:tauandwpoflambda}. We find that the relaxation time diverges and the plasmon frequency vanishes as the Coulomb interaction is taken to zero. The  behavior of $\omega_P$ is in agreement with the weakly coupled logic: it grows with the net charge of the system $\mu/T$ and with the strength of the Coulomb interaction  $\lambda$, vanishing in the limit $\mu/T\rightarrow 0$.

We emphasize that in the system under investigation momentum is conserved, so $\tau$ cannot be thought of as the Drude relaxation time. As already mentioned in the previous section, the effective relaxation scale $\tau$ appearing in the model~\eqref{model} is not derived microscopically and it is understood as an effect of the quantum critical sector of the theory,~\cite{Krikun:2018agd}. It would be desirable to find a more quantitative understanding of the microscopic phenomena governing this effect and its relation to the polarization of the underlying degrees of freedom.

\begin{figure}[t!]
    \centering
   \tcbox{ \includegraphics[width=0.8 \linewidth]{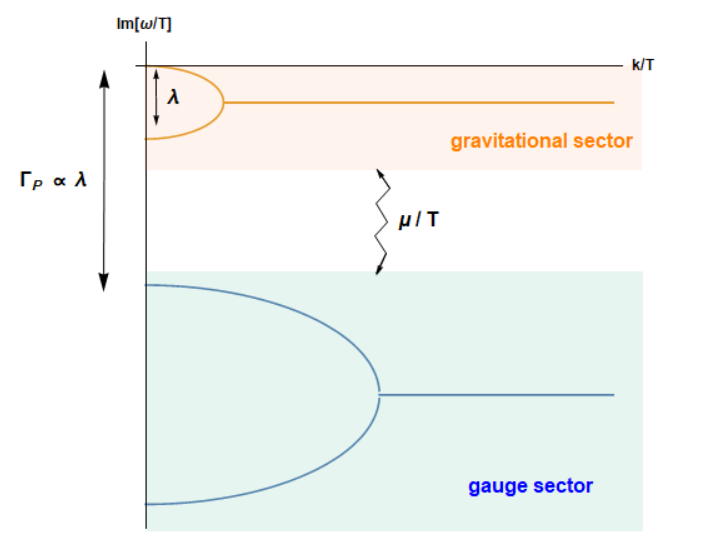}}
    \caption{A simplified summary of the interplay between the gravitational sector and the gauge field sector at finite chemical potential. The dynamics is explained in terms of the chemical potential $\mu$ and the electromagnetic coupling $\lambda$. The zig-zag arrow, mediated by the dimensionless parameter $\mu/T$, represents the coupling between the gauge sector (blue area) and the gravitational sector (orange area). Increasing the chemical potential, the two sectors become more and more interacting.}
    \label{sumfig}
\end{figure}

Despite of the surprising success of the simple analytic model~\eqref{model} in fitting the numerical data, there are several open questions. First, the choice of the model was purely phenomenological. Second, the nature of the relaxation time $\tau$ and the definition of the conductivity~\eqref{cond} is still obscure. Finally, the role of the incoherent conductivity has not been discussed. Our observation is that the introduction of an additional incoherent term $\sigma_0$ in the definition of eq.~\eqref{model} spoils the agreement with the numerical data. 
Before proceeding with the introduction of translational symmetry breaking, let us summarize the dynamics of the modes discussed in this section in a schematic picture (see figure~\ref{sumfig}).

At finite $\lambda$ and zero chemical potential the two sectors are completely decoupled at linear order in the perturbations. The gravitational sector contains a hydrodynamic sound mode and the gauge sector, analyzed in section~\ref{sec:schwarzschild}, displays two damped non-hydrodynamic modes. The most damped one does not depend on $\lambda$ (to first order in momentum) and will not play any relevant role in the final picture. In contrast, the smallest damping, $\Gamma_P$, increases linearly with $\lambda$. As a consequence, for small coupling this damping is very small and it enters in the hydrodynamic window $\omega/T \ll 1$.

Now, let us switch on a small chemical potential $\mu/T \ll 1$. The chemical potential couples the two sectors and, together with finite electromagnetic coupling $\lambda$, produces the transition of the sound mode into a low momenta diffusive mode. This mode then collides with a second non-hydro mode and produces the $k$-gap picture we see in figure~\ref{sumfig}. Moreover, due to conservation of momentum, there is always a hydrodynamic mode in the spectrum. Now, at large $\lambda \gg 1$, the modes stemming from the gauge sector are strongly damped and they do not play any role in the combined dynamics. This is shown in the blue region in figure~\ref{sumfig}. As discussed in the $\mu/T=0$ case, for a small electromagnetic coupling one of the modes of the gauge sector becomes long-living, giving rise to the standard charge diffusion mode in the extreme $\lambda=0$ limit. On the other hand, the modes in the gravitational sector approach each other as $\lambda$ decreases and give rise to the sound mode of RN.
When the chemical potential is large enough, the two sectors are strongly mixed; the first mode from the gauge sector enters the ``hydrodynamic'' regime and causes the interplay between the three complex modes discussed in the text, first observed in~\cite{Gran:2018vdn}. In the limit of very large $\mu/T$ we recover the standard (bulk\footnote{Also called a volume or co-dimension zero plasmon, in contrast to surface or interface plasmons. Hence, `bulk' here should not be confused with the bulk spacetime.}) gapped plasmon dispersion relation $\omega^2\,=\,\omega_P^2\,+\,c^2\,k^2$.

\section{Plasmons with broken translations}\label{sec:breaking}
So far, we have considered a holographic background which enjoys translational invariance. Momentum, as an operator of the dual field theory, is a conserved quantity. The momentum conservation equation produces several effects on transport, including an unnatural divergent contribution to the electric conductivity. In realistic condensed matter systems translation (and rotation) invariance is generally broken because of the presence of impurities, disorder or more complicated mechanisms. In this section, we make use of a generalization of the holographic picture that allows to break translational invariance, retaining the simplicity of the framework~\cite{Vegh:2013sk,Davison:2013jba}. More specifically, we consider the additional scalar sector in action~\eqref{eq:action} and, without loss of generality, set $m^2=1$. The explicit breaking of translational invariance introduces a finite relaxation rate for momentum $\Gamma_M=\tau_M^{-1}$, related to the mass of the graviton fluctuations, which in our case is determined by the dimensionless parameter $\alpha/T$. The stress tensor (or at least its spatial component) is not conserved anymore and the corresponding hydrodynamic modes get strongly affected. There is a vast literature on the topic; for more details we refer to~\cite{Baggioli:2016rdj}.

We make use of the linear axions model of~\cite{Andrade:2013gsa} to relax momentum in the dual field theory. The Ward identity for the momentum operator is broken and at linear order\footnote{The linear order approximation corresponds to the limit of slow momentum dissipation $\tau_M T \gg 1$. From the bulk perspective it corresponds to consider the graviton mass small compared to the temperature of the system $T$.} it reads:
\begin{equation}\label{anal}
    \partial_i\,T^{it}\,=\,0\,,\quad \quad  \partial_i\,T^{ij}\,=\,-\,\Gamma_{M}\,T^{jt}\,+\,\dots
\end{equation}
where the ellipsis indicate higher order corrections relevant for fast momentum dissipation. In the limit of slow momentum relaxation $\Gamma_m/T \ll 1$, the relaxation rate is defined (see~\cite{Davison:2013jba}), by the analytic expression
\begin{equation}
    \Gamma_M\,=\,\frac{s}{4\,\pi\,(\varepsilon+p)}\,\alpha^2\,+\,\mathcal{O}(\alpha^4)\,,\label{relrel}
\end{equation}\\
where $s$, $\varepsilon$ and $p$ are the entropy density, the energy density and the pressure of the system and $\alpha$ the graviton mass\footnote{More precisely the mass of the helicity-1 component of the graviton. Notice that the massive gravity theory considered is not Lorentz invariant and therefore the various components of the graviton can acquire different masses~\cite{Dubovsky:2004sg,Alberte:2015isw}.}.
At this point, we can investigate the effects of momentum dissipation on the dispersion relation of the plasmons. 

 \begin{figure}[t!]
\centering
\includegraphics[width=0.48\linewidth]{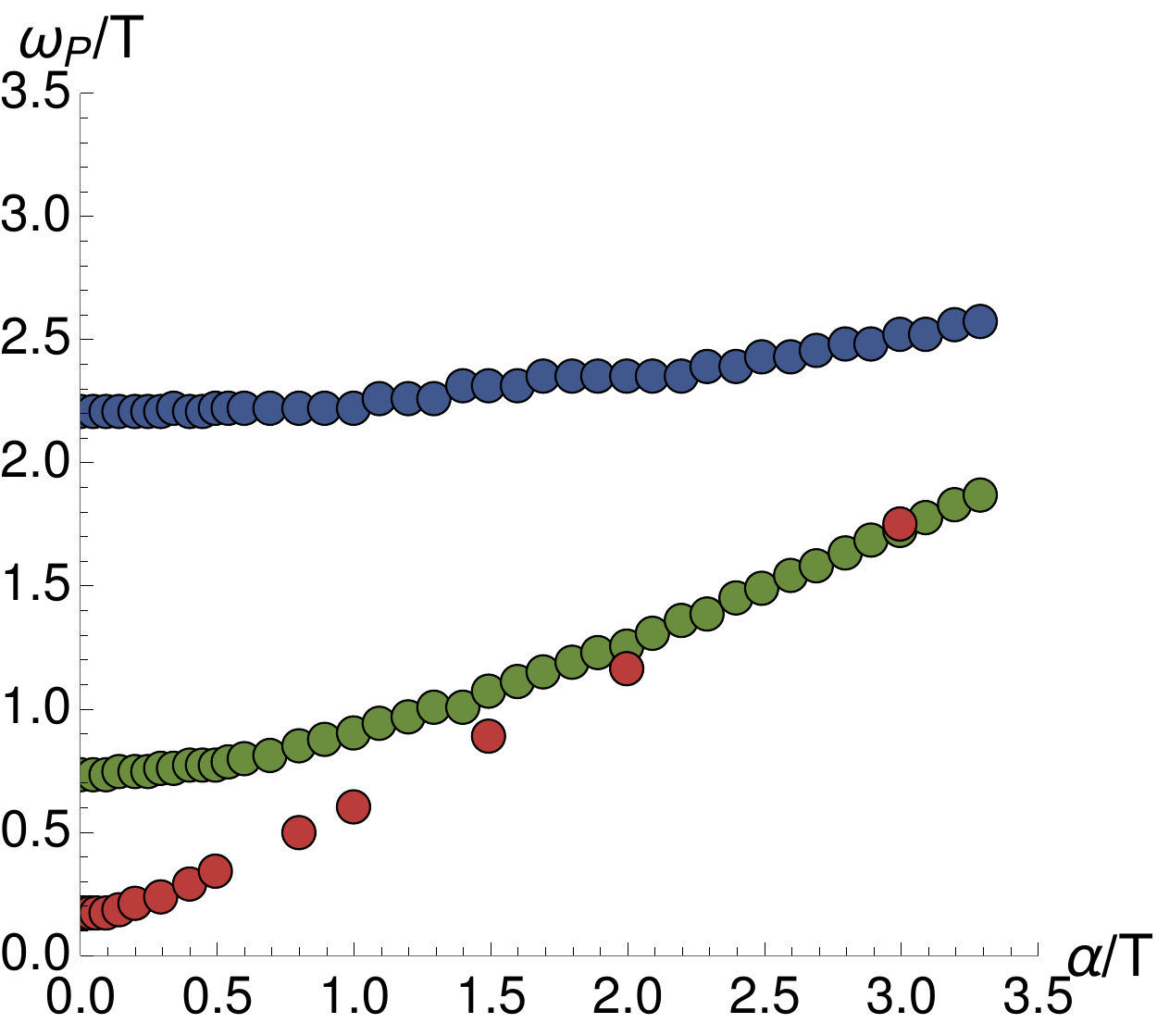}\quad
\includegraphics[width=0.48\linewidth]{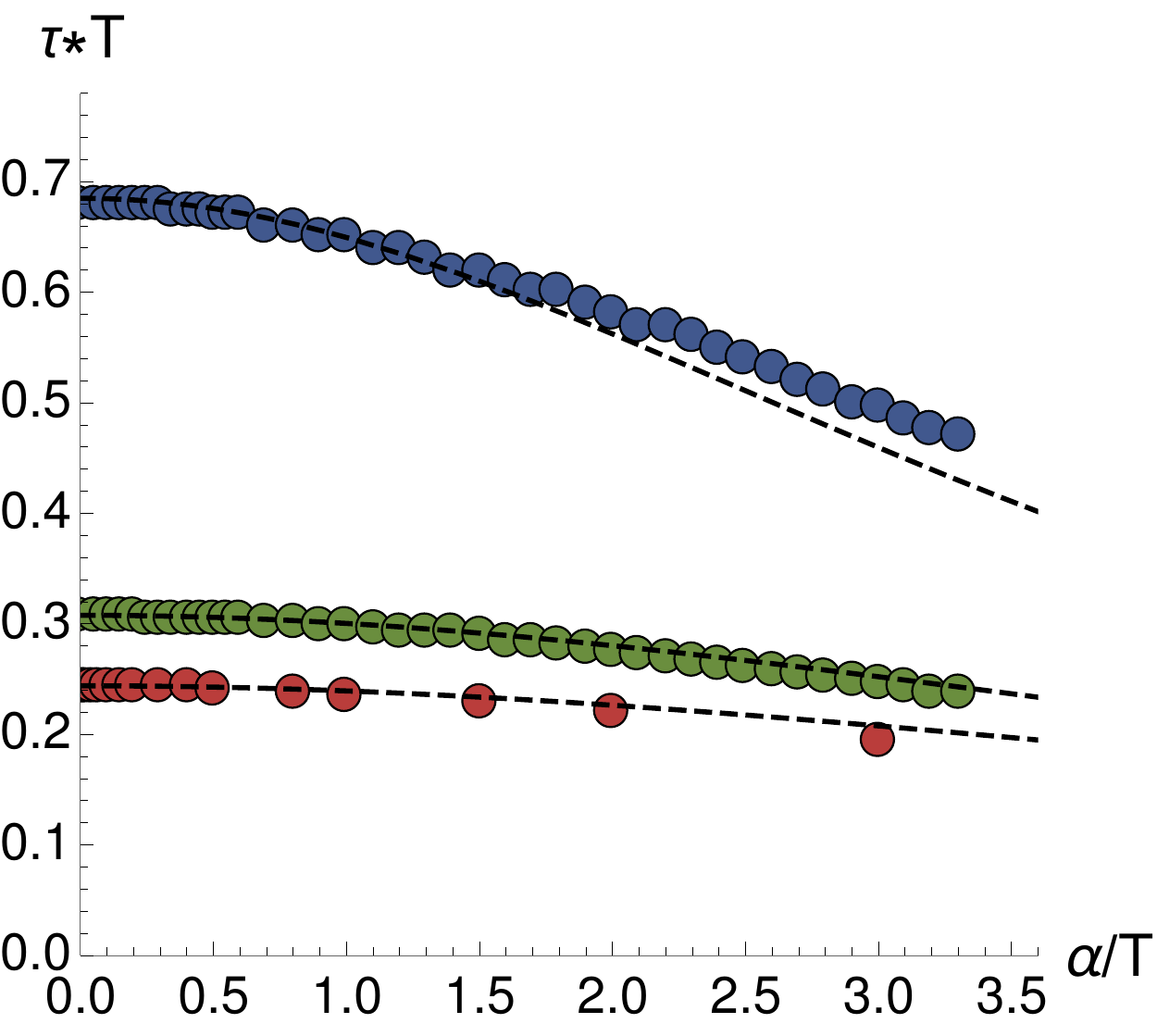}
\caption{Evolution of $\omega_P$ and $\tau$ with increasing $\alpha$ for $\mu/T=[0.2,1,4]$ (red, green, blue) and fixed $\lambda=1$. The dashed lines are the analytic predictions of \eqref{relrel} and~\eqref{matt} at small $\alpha/T \ll1$. The agreement between the data and the analytic correction~\eqref{relrel} is apparent at $\alpha/T \ll 1$.}
\label{fig:tauofalpha}
\end{figure}

Firstly, it is important to remark that we still find the expression obtained in~\eqref{model} to be a good approximation for the dynamics of the three lowest collective modes in the presence of momentum dissipation. The effect of a finite $\alpha/T$ is simply to renormalize the effective parameters entering in eq.~\eqref{model}. The numerical results are shown in figure~\ref{fig:tauofalpha}. Increasing momentum dissipation, the relaxation time entering in eq.~\eqref{model} becomes smaller. This can be simply understood by introducing a new contribution to the relaxation rate via the inverse Matthiessen' rule:
\begin{equation}
    \tau_{total}^{-1}\,=\,\tau_{M}^{-1}+\tau^{-1}~.\label{matt}
\end{equation}
This formula, combined with the momentum relaxation rate defined in~\eqref{relrel}, approximates correctly $\tau_{total}$ at slow momentum dissipation rates $\Gamma_M/T \ll 1$ as shown in figure~\ref{fig:tauofalpha}.

Secondly, we observe that the plasma frequency increases with the momentum dissipation parameter $\alpha/T$. From a weakly coupled perspective, where the plasma frequency is related to the inverse of the effective mass of the charge carriers $m^*$, this result is surprising. It suggests that momentum dissipation decreases the effective mass of the charge carriers, while at weak coupling the more scattering events the heavier the charge carriers' effective mass becomes. However, given the absence of well-defined quasi-particles in the strongly coupled holographic framework arguments using weak coupling are not expected to be applicable. Moreover, we can notice that in the transverse spectrum of the linear axions theory, at $\lambda=0$, an analogous phenomenon happens. More specifically, at very strong momentum dissipation, a mode with a massive particle dispersion relation appears in the spectrum~\cite{Baggioli:2018vfc,Baggioli:2018nnp}.

Finally, let us notice that, even in presence of strong momentum dissipation, \ie$\Gamma_M/T \gg 1$, a hydrodynamic mode is present in the longitudinal spectrum. As already shown in~\cite{Davison:2014lua}, such a mode simply corresponds to the conservation of energy and at very strong momentum relaxation, \ie the usually called \textit{incoherent limit}, it simply encodes the energy diffusion mode.

\section{The effects of the spontaneous breaking of translational invariance}
\label{sec:breaking2}
In this last section we consider a different symmetry breaking pattern; the spontaneous breaking (SSB) of translations. We assume\footnote{As shown in~\cite{Alberte:2017oqx}, this concrete choice of the potential does not affect the generality of the picture. Similar results can be obtained using potentials of the form $V(X)=X^N$ with $N>5/2$.}:
\begin{equation}
    V(X)\,=\,X^3\,,
\end{equation}
and we fix for simplicity $m^2=1$. The nature of this second mechanism is very different. The SSB of translations does not introduce any relaxation mechanism but it does give new and dynamical degrees of freedom, which are identified as the Goldstone bosons for translational invariance, \ie the phonons~\cite{Leutwyler:1996er}. As a physical consequence, the SSB of translations is directly responsible for the elastic properties of the medium~\cite{landau7}. The strength of the SSB is parametrized by the dimensionless parameter $\alpha^3/T$. We will provide more details in the following (see~\cite{Ammon:2019wci} for a recent discussion about the explicit versus the spontaneous breaking of translations in this context).

Importantly, we consider the longitudinal sector of the system and therefore what we discuss is the longitudinal sound\footnote{For a discussion on transverse sound in the same model see~\cite{Alberte:2017oqx,Ammon:2019wci}.}. In contrast to the transverse sound, the longitudinal counterpart is present in both liquids and solids and it usually relates to the compressibility of the material. 
From a very broad perspective, derived in details in \cite{landau7,PhysRevA.6.2401,PhysRevB.22.2514}, the speed of longitudinal sound in generic systems, displaying the spontaneous breaking of translations, is given by:
\begin{equation}
    v_L^2\,=\,\frac{K\,+\,G}{\chi_{PP}}
\end{equation}
where $G$ and $K$ are the shear and bulk elastic moduli, and $\chi_{PP}$ the momentum susceptibility. Just using thermodynamics, the bulk modulus $K$, which is the inverse of the compressibility, can be defined as:
\begin{equation}
    K\,=\,-\,V\,\frac{\partial P}{\partial V}
\end{equation}
where $V$ is the volume and $P$ the mechanical (not the hydrodynamic) pressure, $P\equiv T_{xx}$. Importantly, the total compressibility can divided into two contributions:
\begin{equation}
    K\,=\,\frac{\partial p}{\partial \varepsilon} \,\chi_{PP}\,+\,\kappa
\end{equation}
where this time $p$ is the hydrodynamic pressure $p=P-2\,\kappa$, and $\kappa$ is the contribution to the bulk modulus which comes directly from the SSB of translational invariance.\\
Consider a system where translations are not broken, not even spontaneously. It follows that:
\begin{equation}
    G\,=\,\kappa\,=\,0\,,\quad \quad P\,=\,p
\end{equation}
Therefore, the speed of longitudinal sound drastically simplifies into:
\begin{equation}
    v_L^2\,=\,\frac{\partial p}{\partial \varepsilon}\,,\label{lala}
\end{equation}
in agreement with hydrodynamics~\cite{Kovtun:2012rj,Policastro:2002tn}. Moreover, in a conformal field theory, where $\varepsilon=(d-1) p$, the speed takes the constant value $v_s^2=1/(d-1)$, with $d$ the number of space-time dimensions.
It is illustrative to rewrite the total speed of longitudinal sound as:
\begin{equation}
    v_L^2\,=\,\frac{\partial p}{\partial \varepsilon}\,+\,\frac{\kappa\,+\,G}{\chi_{PP}}\label{elel}
\end{equation}
Defining the spontaneous breaking scale as $\langle SSB \rangle$, the first term is $\sim\langle SSB \rangle^0$, while the second one is $\sim\langle SSB \rangle^2$. See \cite{Ammon:2019apj,oriolfuture,Baggioli:2019abx} for more details.\\
This analysis indicates that, even in the presence of SSB of translational invariance, only one sound mode appears in the longitudinal spectrum and its speed is determined by two different contributions:
\begin{equation}
    v_L^2\,=\,v_0^2\,+\,v_{SSB}^2 \,.\label{total}
\end{equation}
as discussed in~\cite{Delacretaz:2017zxd,Ammon:2019apj,Baggioli:2019abx}.
We will denote the first contribution, $v_0^2$, which is finite in absence of SSB, as ``zero sound''. The naming and the identification are clearly a stretch, because the presence of a Fermi surface in these systems is far from obvious~\cite{Kulaxizi:2008kv,Kulaxizi:2008jx,HoyosBadajoz:2010kd,Davison:2011ek}. Moreover, such sound mode appears also for bosonic holographic systems~\cite{Davison:2011uk}. Additionally, the label ``zero sound'' usually refers to zero temperature; here we extend it to finite, but small, temperatures without introducing any further distinction. All in all, for simplicity, we keep the name. In simple terms, we can think of that contribution as the one coming from the ``fluid'' soup provided by the strongly coupled CFT, the critical continuum of our holographic setup. In our setup, this sector is charged under a U(1), introduced via the Maxwell field of the holographic bulk theory.

The second term in~\eqref{total}, denoted with $v_{SSB}^2$, is given by the elastic response of the material; it is directly related to the existence of longitudinal Goldstone bosons and it vanishes in the limit of no SSB of translations. We will denote for simplicity this second term the ``normal sound'' contribution. This second term is tightly connected to the presence of the scalar sector $\phi^I$. Importantly, the scalar fields are not charged under the $U(1)$ symmetry of the system.
\begin{figure}[t!]
    \centering
    \includegraphics[width=11cm]{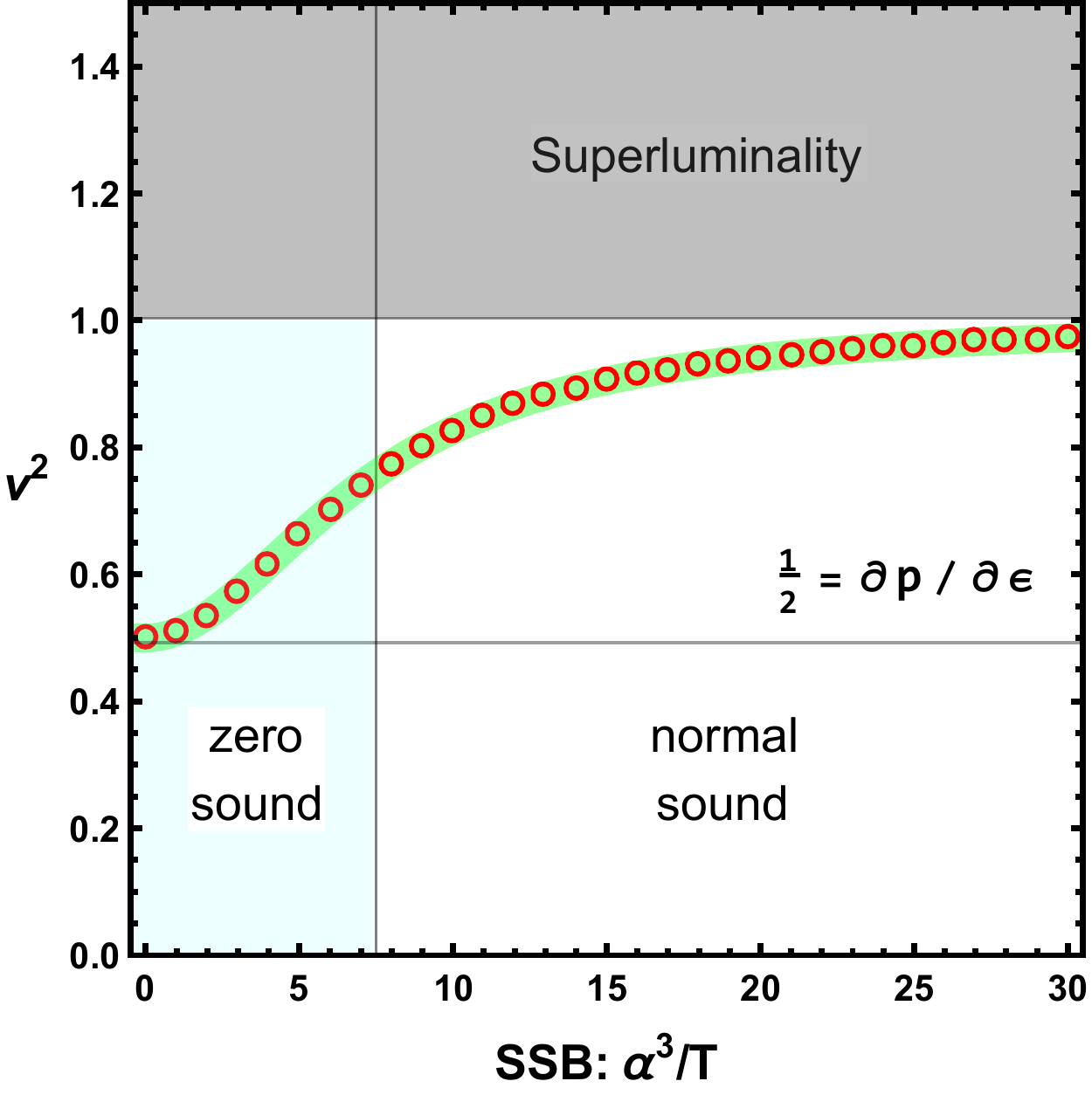}
    \caption{The speed of longitudinal sound as a function of the dimensionless SSB scale $\alpha^3/T$. The red dots are numerical data extracted using formulae \eqref{lala}, \eqref{elel} and \eqref{total}.  The superluminal region $v_L^2 >1$ is displayed in the phase diagram with shaded color. The crossover between two regimes is clear and it happens around $\alpha^3/T \sim 7.5$. At small SSB the speed of sound is close to the zero sound limit $v_0^2= \partial p/\partial \epsilon =1/2$. At higher value of the SSB parameter the speed is mainly governed by the elastic properties of the material; we call that ''normal sound''. For simplicity, this plot is done at $\mu=0$.}
    \label{figvel}
\end{figure}\\
In our holographic setup, we have a dimensionless parameter $\alpha^3/T$ which determines the strength of the SSB of translations. We can now quantify the speed of longitudinal sound as a function of this parameter. The numerical results, supporting these statements, are found in figure~\ref{figvel}. Our results are thus consistent with those from recent preprints~\cite{Ammon:2019apj,Baggioli:2019abx}. The speed of sound interpolates between two  regimes:
\begin{enumerate}
    \item For vanishing or small SSB ($\alpha^3/T \ll 1$), the sound is mainly governed by what we called ``zero sound'' and it is given by the hydrodynamic formula~\eqref{lala}.
    \item For large SSB, on the contrary, the sound is mainly controlled by the 
    ordered structure (``lattice'') and its speed is dominated by the elastic contribution, the second term in eq.~\eqref{elel}.
\end{enumerate}
This mechanism has been already directly observed in~\cite{Andrade:2017cnc}. Let us notice that this transition between a solid like behaviour, where longitudinal sound is controlled by the elastic moduli as in ordered crystals, and a fluid type regime where sound behaves more like the electron sound of a Fermi liquid, is also evident in several other observables. We can mention for example the shear elastic modulus and the dynamics of transverse shear waves analyzed in~\cite{Alberte:2017oqx,Ammon:2019wci}. In the limit of large temperatures, compared to the graviton mass governing the SSB, the system loses the elastic properties in a continuous way similar to a glass transition. This observation was already present in several previous works in the literature~\cite{Andrade:2017cnc,Alberte:2017cch,Alberte:2017oqx,Baggioli:2018bfa,Andrade:2019zey}.

 \begin{figure}[t!]
\centering
\includegraphics[width=0.48\linewidth]{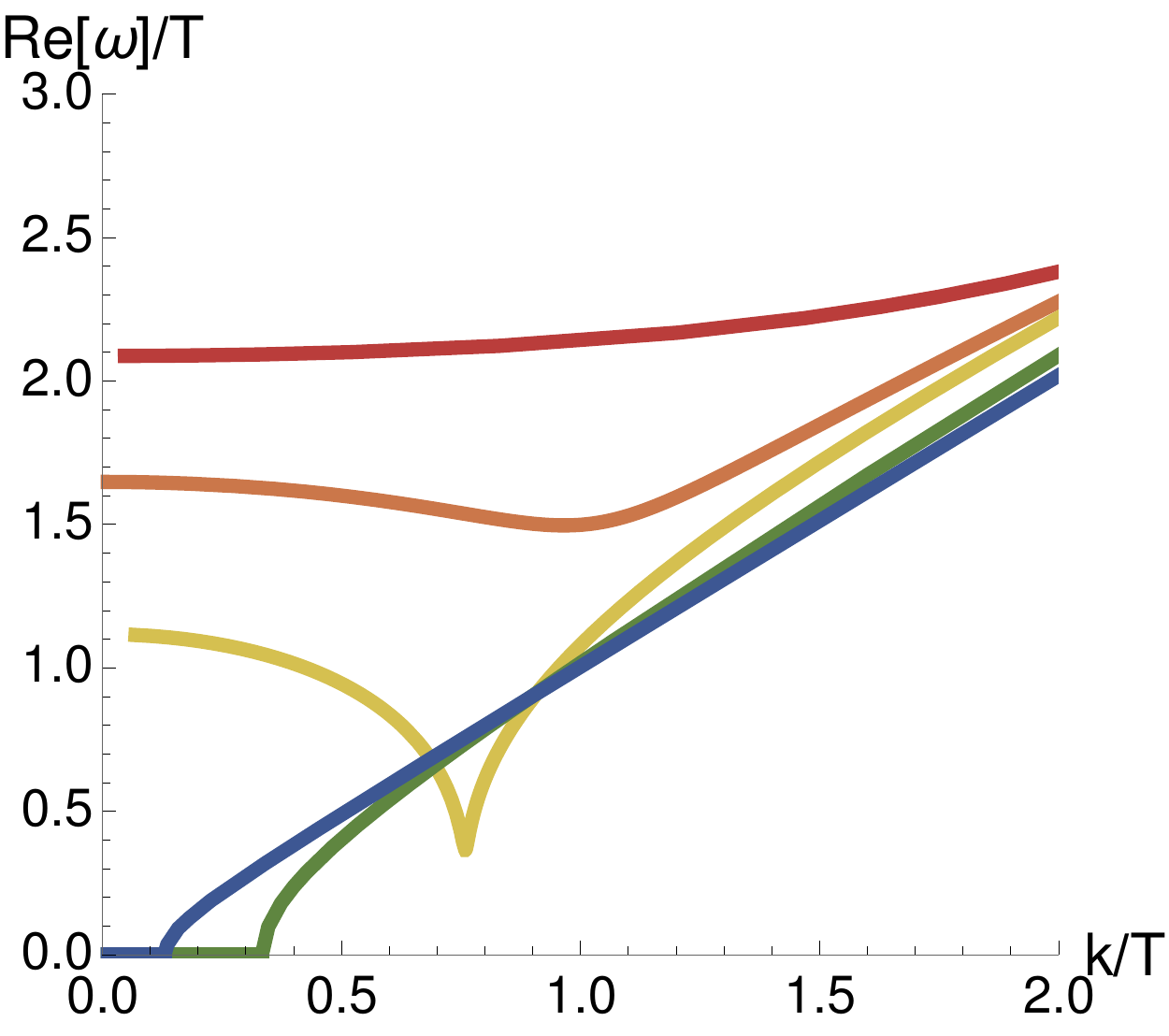}\quad
\includegraphics[width=0.48\linewidth]{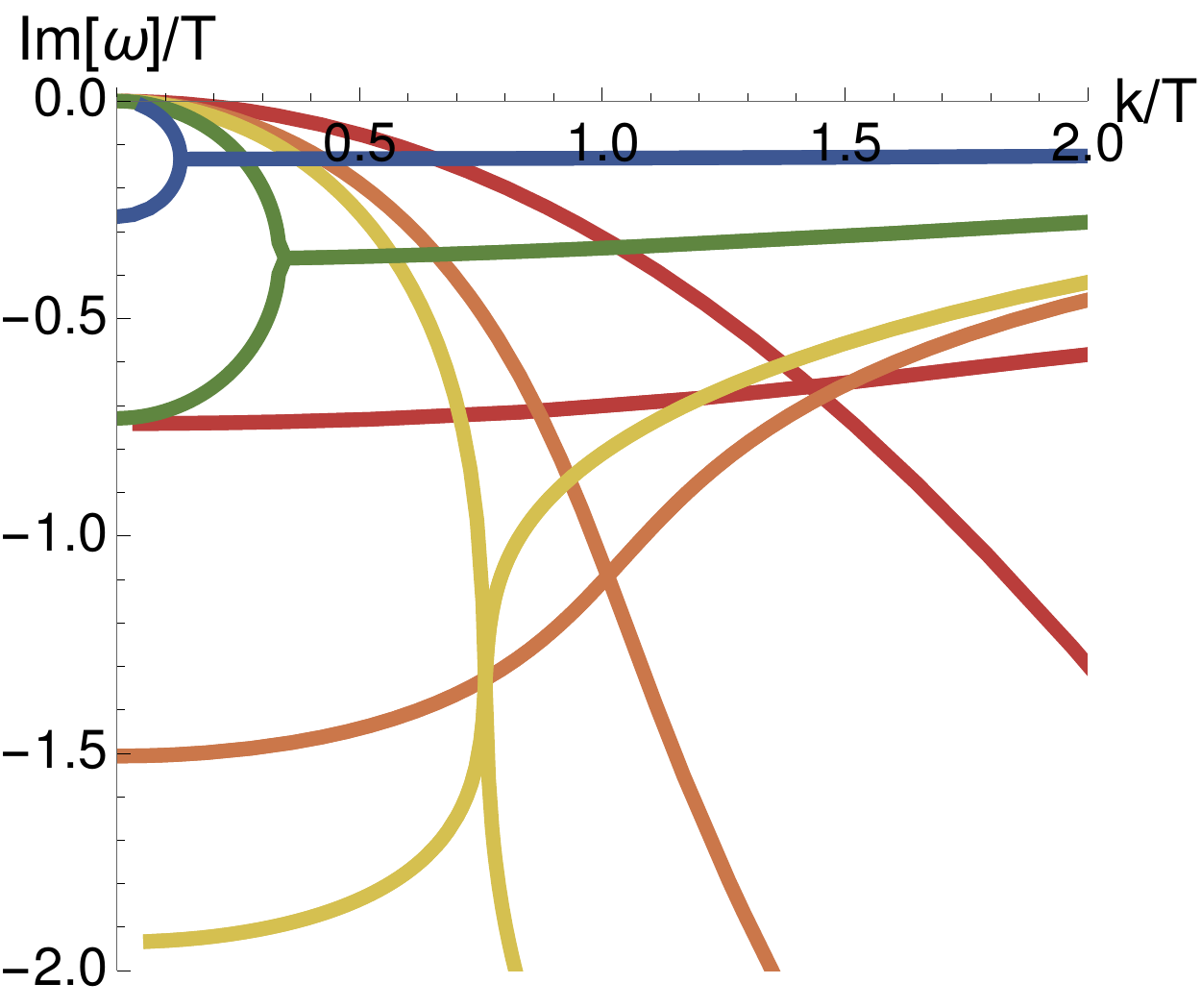}\\[0.5cm]
\centering
\includegraphics[width=0.48\linewidth]{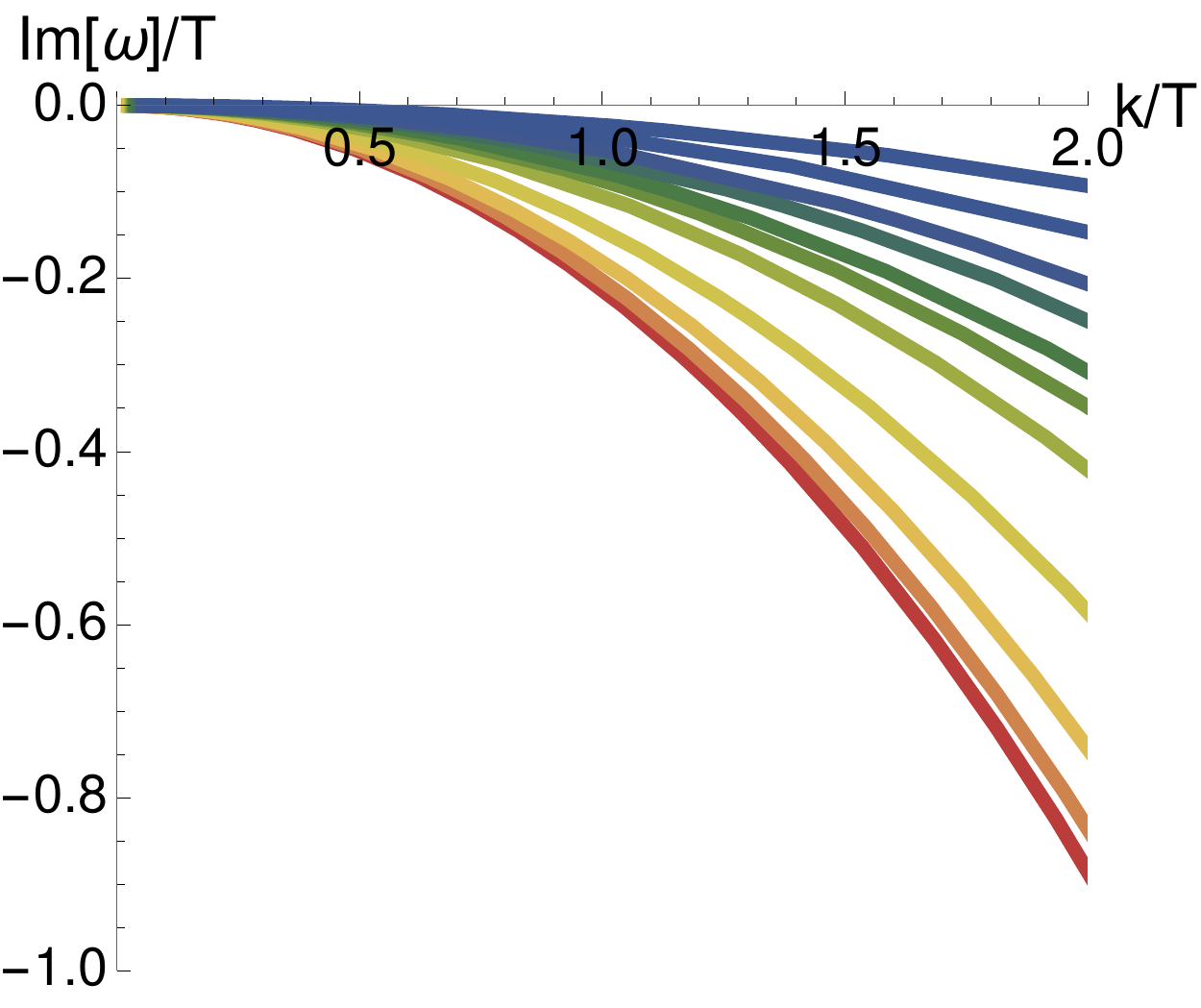}
\caption{\textbf{Top panel: }Real and imaginary parts of the lowest collective modes for fixed $\lambda=1$ and $\mu/T=4$ increasing the dimensionless SSB strength $\alpha^3/T=(0.1,3.75,5,10,25) $ (red-blue). The gapped plasmon smoothly crossovers to the ungapped normal sound mode. For clarity, we manually removed the decoupled crystal diffusion mode. \textbf{Bottom panel:} the decoupled crystal diffusion mode for $\alpha^3/T=[0.1-25] $ (red-blue).}
\label{fig:phonplasmon}
\end{figure}

Before switching on the Coulomb interactions, i.e.~letting $\lambda \neq 0$, let us just point the reader to the known results in absence of them. The analysis of the hydrodynamic modes can be found in standard textbooks such as \cite{Lubensky,landau7} or in more modern works such as \cite{Delacretaz:2017zxd} and it was recently tested in holography in \cite{Ammon:2019apj}.

Now, we study the behaviour of the system when long range Coulomb interactions are taken into account. A comment regarding our setup is in order here. In some cases in the literature \cite{Amoretti:2018tzw}, this kind of models have been referred to as Wigner Crystals. Here we argue that there is no concrete indication that this is the case, at least in our particular model. Let us remark that what we mean here with Wigner Crystal is the canonical scenario of a lattice of strongly coupled electrons typical of $2D$ systems at very low density and temperature \cite{PhysRev.46.1002,RevModPhys.71.87}. On the contrary, our system is to be understood as follows. The \textit{neutral} scalar sector, which sets the elastic properties of the material, encodes the phononic degrees of freedom, \textit{i.e.} the Goldstone modes for the breaking of translational invariance. One can think here of the neutral, acoustic, phonons of a regular rigid structure. Notice that this is an effective field theory (EFT) description of these modes which does neither attempt to describe any microscopic feature of the underlying lattice nor its dynamical formation. On top of that, the charge sector, encoded in the Maxwell fields in the bulk, is coupled to the phonon sector only indirectly via gravity. The long range Coulomb interactions are implemented in the charged sector via modified boundary conditions and they do not affect directly the neutral scalar sector. Importantly, the effect of the charged sector on the system is mediated by the chemical potential $\mu$ and the strength of the Coulomb interactions $\lambda$, while the effect of the neutral sector is governed by the SSB strength $\sim \alpha$.  Let us remark that, as we show below, our system has only one mode stemming from the hybridization of the would-be plasmon of the charged sector and the phonon, in contrast to the weakly coupled picture.  Qualitatively, our system is more similar to the models used to address the problem of acoustic polarons~\cite{sumi, fantoni}.
This is further supported by our results as we will argue now. 

We investigate the low temperature regime $T/\mu \ll 1$ as we increase the strength of the SSB,  $\alpha^3/\mu$. We already know, from the previous discussion, that in absence of SSB the ``zero sound'' mode gets gapped at low temperature and it shows the typical plasmonic dispersion relation:
\begin{equation}
    \omega^2\,=\,\omega_P^2\,+\,c^2\,k^2\,.\label{io}
\end{equation}
The main question we address here is how the previous dispersion relation is modified in presence of SSB of translations. The results are shown in figure~\ref{fig:phonplasmon}. Increasing continuously the strength of SSB, the gap of the plasmon mode in eq.~\eqref{io} decreases and the dispersion relation eq.~\eqref{io} becomes the one of normal, gapless, sound. This is consistent with our picture; the longitudinal sound (phonons) originating from the long range, uncharged, order structure. Within our model, there is a competition between the effects of the spontaneous symmetry breaking and the electromagnetic interactions. The first tends to keep the sound mode gapless and propagating, while the second tends to gap this mode and transform it into a plasmon. It would be interesting to compare to the situation in which the ``holographic lattice'' is directly charged under the U(1) in the bulk, (as in the case of \cite{Aprile:2014aja}).

Finally, let us notice the SSB of translational invariance introduces a new additional diffusive mode in the longitudinal sector, known as ``crystal diffusion''~\cite{Delacretaz:2017zxd}. This mode has already appeared in~\cite{Andrade:2017cnc} and it is evident in our spectrum as well. It does not couple to the other modes and it is therefore not qualitatively influenced by the electromagnetic coupling $\lambda$; it depends on the SSB strength as shown in the bottom panel of figure~\ref{fig:phonplasmon}.
\begin{figure}[t!]
\centering
\includegraphics[width=0.48\linewidth]{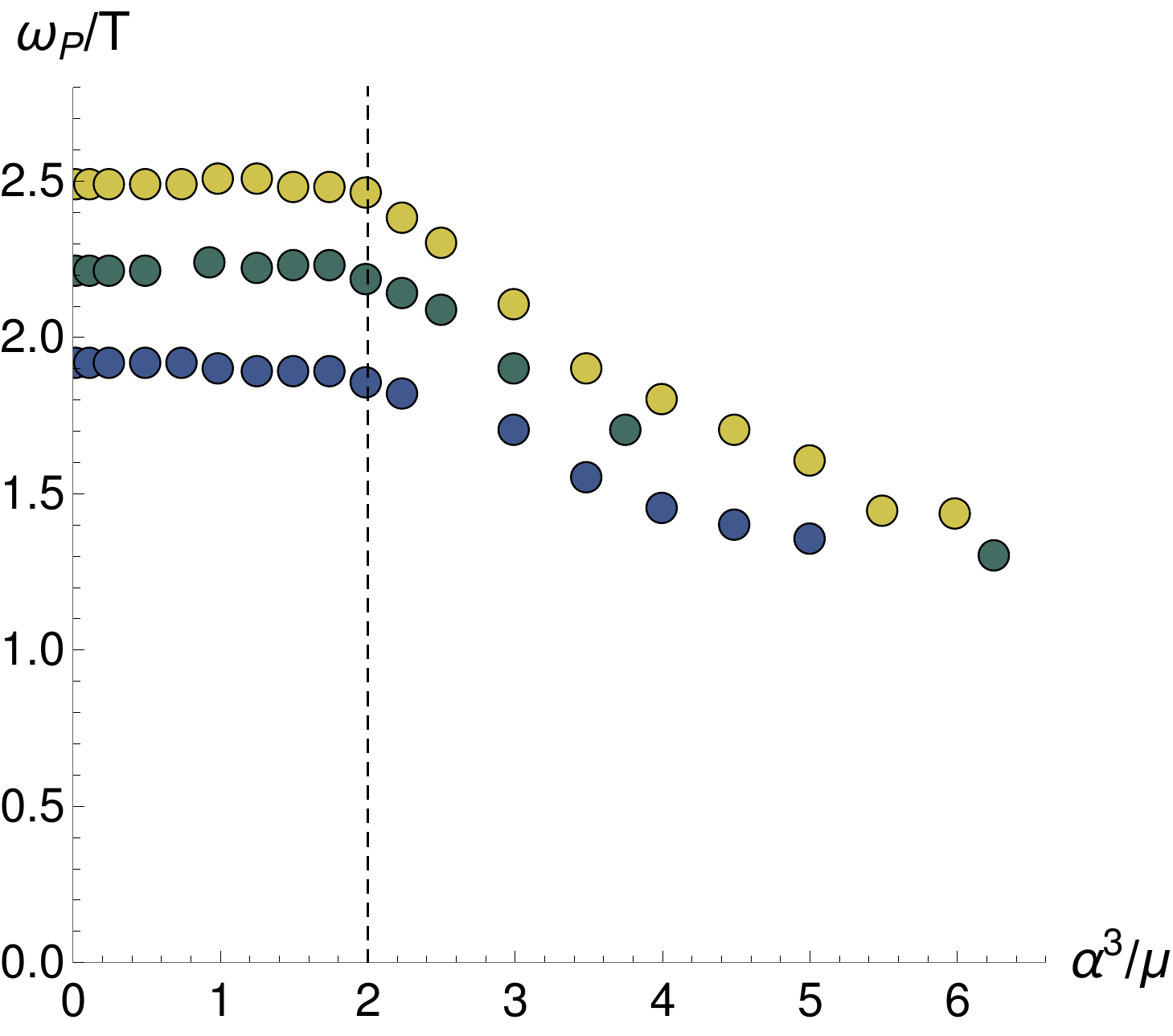}\quad
\includegraphics[width=0.48\linewidth]{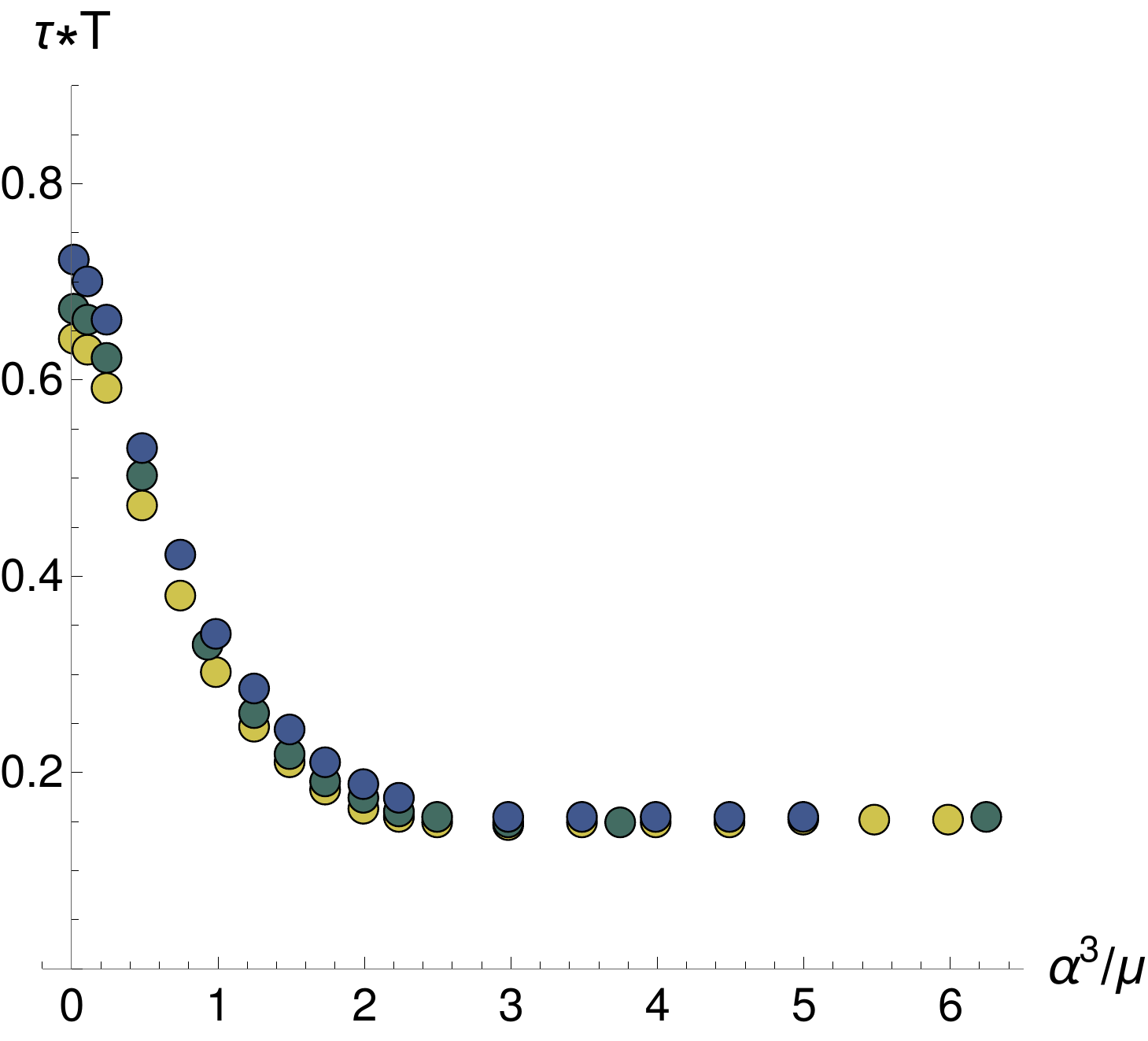}
\caption{The effective relaxation scale $\tau$ and the plasma frequency $\omega_p$ in function of the dimensionless SSB strength $\alpha^3/\mu$ for $\lambda={0.8,1,1.2}$ (blue-yellow). All points are at $\mu/T=4$.}
\label{fig:last}
\end{figure}

\begin{figure}
    \centering
\includegraphics[width=0.34\linewidth]{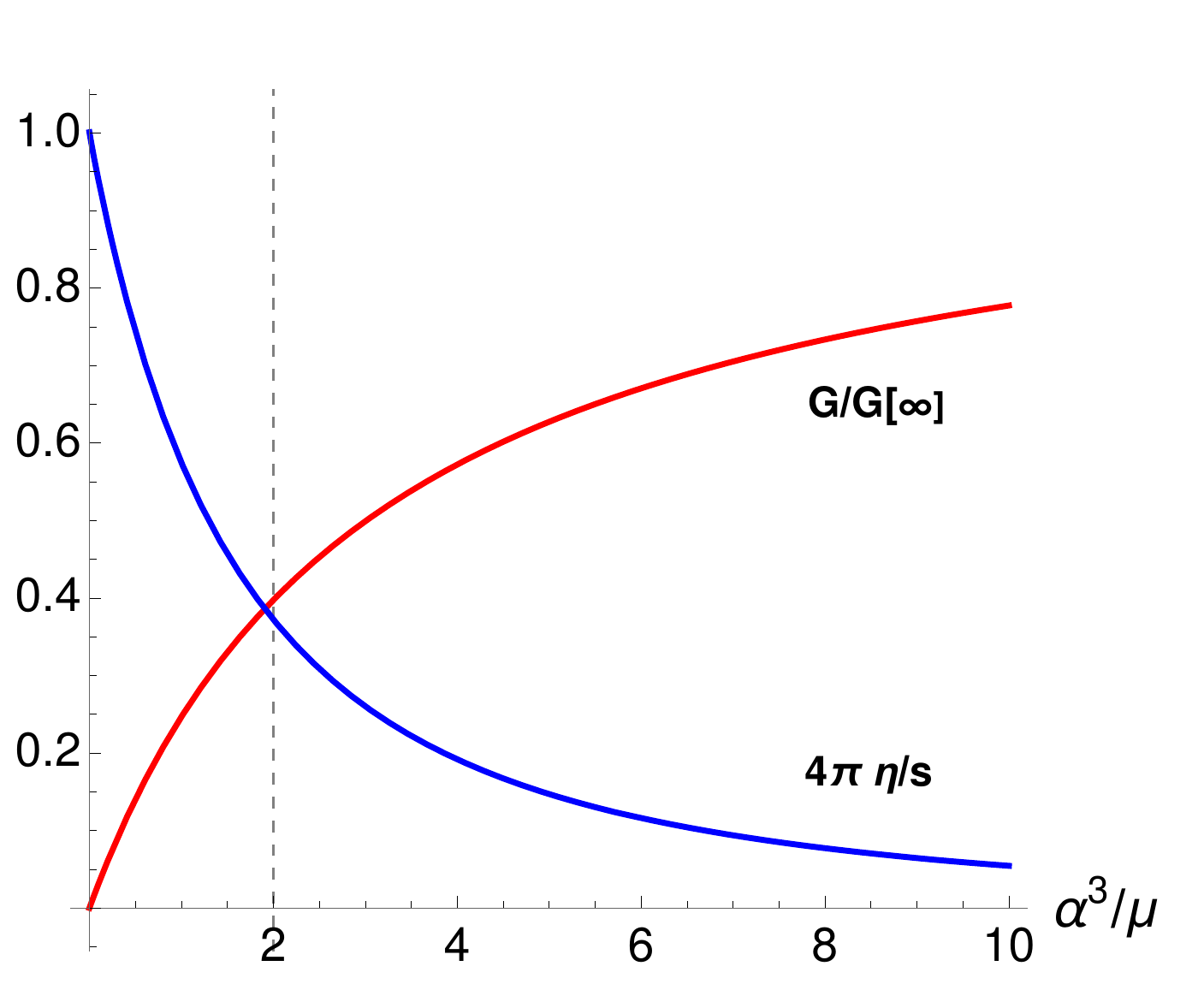}
\quad
\includegraphics[width=0.29\linewidth]{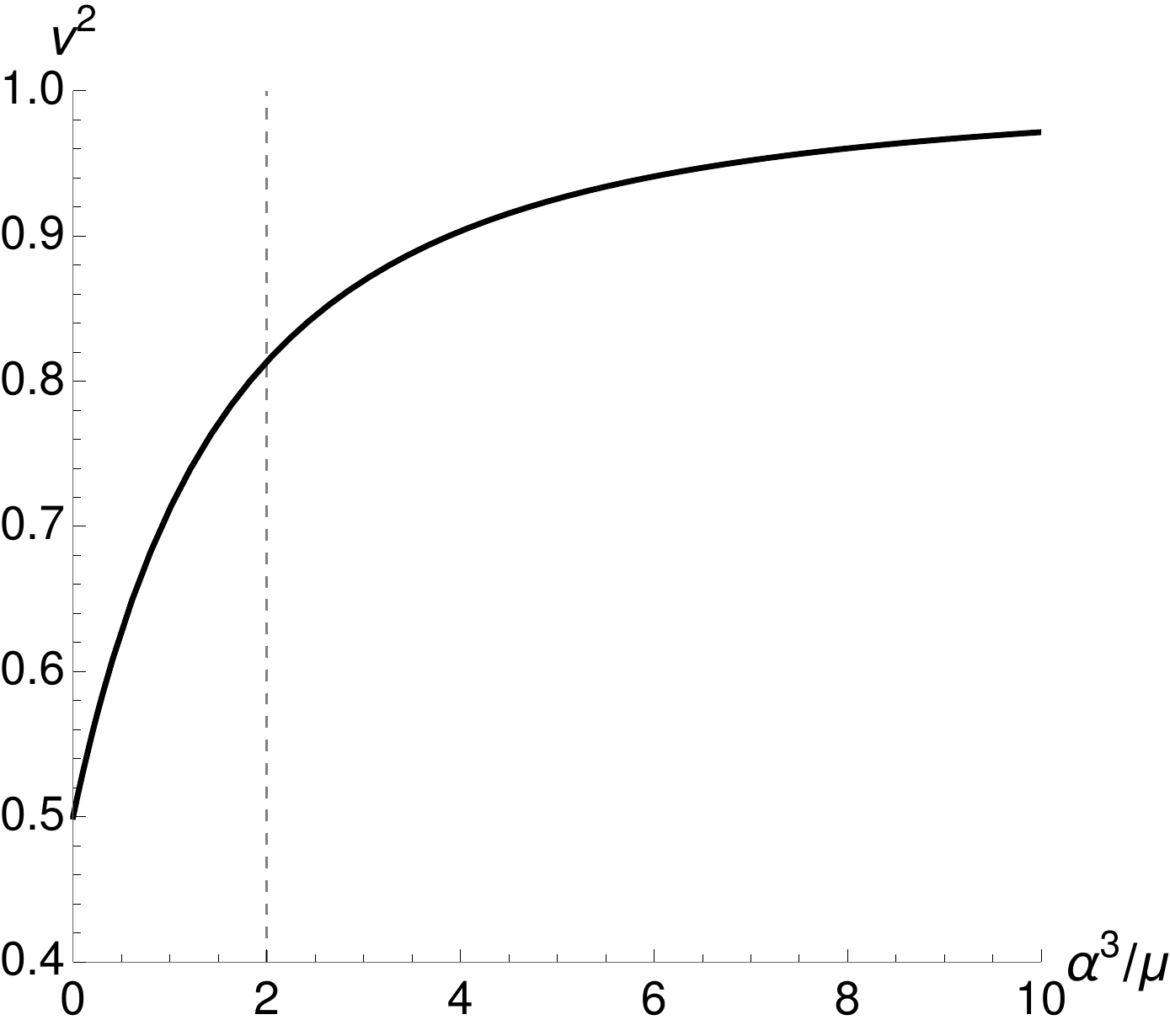}
\quad
\includegraphics[width=0.29\linewidth]{omepalphamu_withline.pdf}
    \caption{Transition point in various observables. The values appear to be very correlated to each other and independent of $\lambda$.}
    \label{fig:my_label}
\end{figure}

In order to make this discussion more quantitative, we plot in figure~\ref{fig:last} the dependence of the plasma frequency $\omega_P$ and the effective relaxation scale $\tau$ as functions of the dimensionless SSB parameter $\alpha^3/\mu$ by fitting the data presented in figure~\ref{fig:phonplasmon}. We observe that before a certain ``critical'' $\alpha^3/\mu$ is reached, the plasma frequency remains constant and independent on the elastic moduli. After this, the plasma frequency sharply decreases, by increasing the amount of SSB as discussed previously. Only in the low $\alpha^3/\mu$ limit, where $\omega_P/T$ is independent of $\alpha^3/\mu$, can it be thought of as the sound of a charged fluid. Notice that the critical $\alpha^3/\mu$ value at which the plasma frequency starts to decrease is independent of the electromagnetic coupling $\lambda$ and in this sense it is rather connected with the property of the background. Interestingly, the transition in the plasma frequency appears to be much sharper than in other observables, see fig \ref{fig:my_label}. In particular, we numerically observe a robust correlation between:
\begin{itemize}
    \item[(I)] The value of $\alpha^3/\mu$ at which the dissipative fluids effects (related to the viscosity $\eta$) and the elastic effects (governed by the shear modulus $G$) become comparable. This point qualitatively signals the transition from a viscous fluid to a rigid material.
    \item[(II)] The crossover value of the longitudinal speed of sound between its $\alpha=0$ value $v_L^2=1/2$ typical of a conformal fluid and its maximum value $v_L^{max}$ obtained in the limit of $\alpha^3/\mu \rightarrow \infty$ where dissipation becomes subleading and the system a perfect solid.
    \item[(III)] The critical value at which the plasma frequency stops to be constant and starts to decrease rapidly in function of $\alpha^3/\mu$.
\end{itemize}
This correlation is highlighted in figure~\ref{fig:my_label} and it confirms our interpretation of the model at hand.
Furthermore, the effective relaxation time $\tau$ seems to decrease with the translational symmetry breaking as well. This feature is slightly counter-intuitive because one does not expect the SSB of translations to contribute to the relaxation of the system. Anyway, let us first remind the reader that the relaxation scale is completely phenomenological and has no direct microscopic interpretation. Moreover, let us notice that in the deep SSB regime, at large $\alpha^3/\mu$, where the physics is more and more that of an ordered crystal\footnote{Despite translations are broken no characteristic lattice length scale is present in this class of models. The background metric remains homogeneous. In this sense, these models, are more similar to amorphous solids or glasses rather than standard crystals with a long-range ordered structure. This was empirically confirmed in~\cite{Baggioli:2018qwu}.}, the relaxation time $\tau$ becomes indeed independent of the SSB breaking scale as expected. At the same time, in the same regime, the plasma frequency drops very fast. That said, this is a valuable direction to get a better understanding of these phenomena.

\section{Conclusions}
In this work we study the effect of the electromagnetic interactions on the density response of simple holographic bottom-up models with and without broken translational invariance. We analyze the dispersion relations of the collective longitudinal modes and their dynamics. We introduce the dynamics of the gauge field and Coulomb interactions using a modified set of boundary conditions~\cite{Aronsson:2017dgf}, corresponding to a double trace deformation in the dual field theory~\cite{Witten:2003ya, Mueck:2002gm, Mauri:2018pzq, Krikun:2018agd}. We focus on the relaxation mechanisms affecting the plasmon modes, relevant for the study of quantum critical and strongly coupled materials.

We review the dynamics of a Maxwell gauge field on the uncharged Schwarzschild background in presence of electromagnetic interactions in the dual theory. There, we observe the appearance of a relaxation time scale proportional to the electromagnetic coupling. Our results suggest this relaxation mechanism should be related to the polarization of the holographic medium and it could be thought of as a quantum critical correction to Debye-type relaxation. It is tempting to associate it to the breaking of the global EM symmetry of~\cite{Grozdanov:2018fic}. The dynamics of the associated mode is indeed in agreement with the quasi-hydrodynamic predictions of~\cite{Grozdanov:2018fic} and the $k$-gap phenomenon of~\cite{Baggioli:2019jcm}.
The connection between double trace deformations~\cite{Marolf:2006nd}, the global symmetries picture of~\cite{Grozdanov:2016tdf} and the effects of polarization and screening surely deserves further investigation.

At finite charge density, and linear order in the perturbations, the gauge field couples to the gravitational sector and the dynamics becomes much richer. We discuss a simple phenomenological model~\eqref{model}, which successfully describes the qualitative dynamics of the lowest three modes. Using such a framework, we find that both the plasma frequency and the total effective relaxation time increase with the dimensionless parameter $\mu/T$.

We extend the previous results in the literature~\cite{Aronsson:2017dgf,Gran:2018vdn} by introducing the breaking of translational invariance. This makes the holographic models closer to realistic experimental setups. We consider the introduction of momentum dissipation via a simple effective model known as the linear axions model~\cite{Andrade:2013gsa}. The numerical results confirm that momentum dissipation decreases the relaxation time of the collective modes. In other words, the damping of those modes becomes larger and larger the faster momentum is dissipated, which agrees with the naive intuition. Concretely, we find that the two contributions add up satisfying an inverse Matthiessen's rule:
\begin{equation}
    \tau_{tot}^{-1}\,=\,\underbrace{\tau^{-1}}_{\text{EM interactions}}\,+\,\underbrace{\tau_M^{-1}}_{\text{momentum dissipation}}\,.
\end{equation}
Moreover, we check that for slow momentum dissipation the relaxation time follows indeed the analytic formulas derived previously in the literature~\cite{Davison:2013jba} and obtained using hydrodynamics. We find that the plasma frequency increases with the momentum dissipation rate.

Finally, we analyze the longitudinal collective modes of the system in presence of the SSB of translational invariance. The qualitative results are very interesting. First, we confirm the appearance of the crystal diffusion mode already observed and discussed in the literature~\cite{PhysRevA.6.2401,PhysRevB.22.2514}. More importantly, we identify a crossover between a fluid regime at small SSB and a ``crystal'' regime at large SSB. In the first limit, the sound propagation is driven by the strongly coupled CFT collective fluid. In this regime the plasma frequency $\omega_P$, obtained from the fit to the formula \eqref{model}, is independent of the parameter $\alpha^3/\mu$. In the second case, however, the sound is given by the elastic moduli of the system and we find that $\omega_P$ decrease with $\alpha^3/\mu$. In other words, we find a competition between the SSB of translations and the formation of the gap due to long range Coulomb interactions. At infinite $\alpha^3/\mu$, the gap of the sound mode vanishes and the dispersion relation is that of the normal longitudinal sound of systems with spontaneous long-range ordered structures. In that limit, sound is just given by the vibrational modes of the ``lattice'' and its dynamics is controlled by the elastic properties of the latter. This sound mode effectively does not get affected by electromagnetic interactions and it does not acquire any plasma gap. Our results highlight the qualitative difference between this system and a Wigner Crystal.

Despite the amount of papers on the topic, several important questions have not yet been addressed. An interesting future direction would be to understand the role of our results in relation to the recent experiments performed in strange metals and quantum critical materials~\cite{Mitrano5392,husain2019crossover}. Finally, similar dispersion relations have been discussed in the hydrodynamic description of collective modes in Weyl semimetals~\cite{Gorbar:2018nmg}. A $k$-gap dispersion relation for plasmons in type II Dirac semimetals has been suggested in~\cite{2019arXiv190410137S} and it is possibly measurable. Similar results have also been discussed and experimentally observed in the transverse spectrum of Yukawa fluids~\cite{PhysRevLett.97.115001,doi:10.1063/1.5088141}. Our results can clearly be of help in this direction. We plan to come back to these issues in the near future.

\section*{Acknowledgements}
We thank Daniel Arean, Aurelio Bermudez, Oscar Henriksson, Niko Jokela, Sasha Krikun, Karl Landsteiner, Koenraad Schalm,  Jan Zaanen and Alessio Zaccone for several discussions and helfpul comments. M.B. thanks Martin Ammon, Sean Gray, Sebastian Grieninger, Victor Castillo and Oriol Pujolas for their collaboration on related projects. M.B. acknowledges the support of the Spanish MINECO’s “Centro de Excelencia Severo Ochoa” Programme under grant SEV-2012-0249. M.B. thanks NORDITA for the hospitality during important stages of this project. 
U.G. and M.T. are supported by the Swedish Research Council.
A.J. is supported by the ``Atracci\'on del Talento'' program (Comunidad de Madrid) under grant 2017-T1/TIC-5258.
\appendix
\section{Equations of motion}
\label{app1}
We consider the longitudinal fluctuations defined by the following set of fields:
\begin{equation}
    \{\,\text{$\delta $g}_{\text{tt}}\,,\text{$\delta $g}_{\text{tx}}\,,\text{$\delta $g}_{\text{xx}}\,,\text{$\delta $g}_{\text{yy}}\,,\delta A_t\,,\delta A_x\,,\delta \phi_x\,\}\,,
\end{equation}
and we assume the radial gauge $\delta A_z=\delta g_{iz}=0$. In the following we provide the equations of motion, given the model and the ansatz discussed in section~\ref{secmodel}.
\subsection{Linear axions}
For the linear axions potential $V(X)=X$ of section~\ref{sec:breaking} we have:
\begin{flalign}
    \frac{k \omega  \text{$\delta $g}_{\text{tx}}}{f^2}
    +\frac{\text{$\delta $g}_{\text{tt}} \left(-2 z f'+6 f+4 z^4\left(h'\right)^2-k^2 z^2+\alpha ^2 z^2-6\right)}{2 f z^2}\nonumber\\
   +\frac{\text{$\delta $g}_{\text{yy}} \left(-12 f^2+f \left(-2 z^2 f''+8 z f'+2 z^4 \left(h'\right)^2+2 k^2
   z^2+\alpha ^2 z^2+12\right)+2 \omega ^2 z^2\right)}{4 f^2 z^2}\nonumber\\
   +\frac{\text{$\delta $g}_{\text{xx}} \left(-12 f^2+f\left(-2 z^2 f''+8 z f'+2 z^4 \left(h'\right)^2+\alpha ^2 z^2+12\right)
   +2 \omega ^2 z^2\right)}{4 f^2 z^2}\nonumber\\
    +\left(\frac{3 f'}{2 f}-\frac{2}{z}\right) \text{$\delta$g}_{\text{tt}}'
    +\frac{f' \text{$\delta $g}_{\text{xx}}'}{4 f}
    +\frac{f' \text{$\delta $g}_{\text{yy}}'}{4f}
   -\frac{3 z^2 h' \text{$\delta $A}_t'}{f}-\frac{i \alpha  k \delta \phi _x}{2 f}+\text{$\delta $g}_{\text{tt}}''&=0\,,
\end{flalign}
\begin{flalign}
    \frac{\text{$\delta $g}_{\text{tx}} \left(-2 z f'+6 f+z^4 \left(h'\right)^2-6\right)}{f z^2}
    +\frac{k \omega \text{$\delta $g}_{\text{yy}}}{f}
    -\frac{i \alpha  \omega  \delta \phi _x}{f}\nonumber\\
    +2 z^2 h' \text{$\delta
   $A}_x'+\text{$\delta $g}_{\text{tx}}''-\frac{2 \text{$\delta $g}_{\text{tx}}'}{z}&=0,
\end{flalign}
\begin{flalign}
\frac{k \omega  \text{$\delta $g}_{\text{tx}}}{f^2}
-\frac{\text{$\delta $g}_{\text{tt}} \left(-2 z f'+6 f+2 z^4\left(h'\right)^2+k^2 z^2+\alpha ^2 z^2-6\right)}{2 f z^2}\nonumber\\
   +\frac{\text{$\delta $g}_{\text{yy}} \left(-12 f^2+f\left(-2 z^2 f''+8 z f'+2 z^4 \left(h'\right)^2-2 k^2 z^2+\alpha ^2 z^2+12\right)-2 \omega ^2 z^2\right)}{4 f^2z^2}\nonumber\\
   +\frac{\text{$\delta $g}_{\text{xx}} \left(12 f^2-f \left(-2 z^2 f''+8 z f'+2 z^4 \left(h'\right)^2
   +3 \alpha ^2z^2+12\right)+2 \omega ^2 z^2\right)}{4 f^2 z^2}\nonumber\\
   +\frac{3 i \alpha  k \delta\phi _x}{2 f}
   +\left(\frac{3 f'}{4 f}-\frac{2}{z}\right) \text{$\delta$g}_{\text{xx}}'
   -\frac{f' \text{$\delta $g}_{\text{yy}}'}{4 f}
   +\text{$\delta $g}_{\text{xx}}''&=0\,,
\end{flalign}
\begin{flalign}
   -\frac{k \omega  \text{$\delta $g}_{\text{tx}}}{f^2}
   +\frac{\text{$\delta $g}_{\text{tt}} \left(2 z f'-6 f-2 z^4\left(h'\right)^2+k^2 z^2-\alpha ^2 z^2+6\right)}{2 f z^2}\nonumber\\
   +\frac{\text{$\delta $g}_{\text{yy}}\left(12 f^2-f \left(-2 z^2 f''+8 z f'+2 z^4 \left(h'\right)^2+2 k^2 z^2+3 \alpha ^2 z^2+12\right)+2 \omega ^2z^2\right)}{4 f^2 z^2}\nonumber\\
   +\frac{\text{$\delta $g}_{\text{xx}} \left(-12 f^2+f \left(-2 z^2 f''+8 z f'+2 z^4\left(h'\right)^2+\alpha ^2 z^2+12\right)-2 \omega ^2 z^2\right)}{4 f^2 z^2}\nonumber\\
   +\frac{z^2 h' \text{$\delta$A}_t'}{f}
   -\frac{i \alpha  k \delta \phi _x}{2 f}
   -\frac{f' \text{$\delta $g}_{\text{xx}}'}{4f}
   +\left(\frac{3 f'}{4 f}-\frac{2}{z}\right) \text{$\delta $g}_{\text{yy}}'
   +\text{$\delta $g}_{\text{yy}}''&=0\,,
\end{flalign}
\begin{eqnarray}
   -\frac{k^2 \text{$\delta $A}_t}{f}-\frac{k \omega  \text{$\delta $A}_x}{f}-\frac{1}{2} h' \text{$\delta
   $g}_{\text{tt}}'+\frac{1}{2} h' \text{$\delta $g}_{\text{xx}}'+\frac{1}{2} h' \text{$\delta
   $g}_{\text{yy}}'+\text{$\delta $A}_t''=0\,,
\end{eqnarray}
\begin{eqnarray}
   \frac{k \omega  \text{$\delta $A}_t}{f^2}+\frac{\omega ^2 \text{$\delta $A}_x}{f^2}+\frac{f' \text{$\delta
   $A}_x'}{f}+\frac{h' \text{$\delta $g}_{\text{tx}}'}{f}+\text{$\delta $A}_x''=0\,,
\end{eqnarray}
and
\begin{flalign}
   \frac{\delta \phi _x \left(\omega ^2-f k^2\right)}{f^2}
   -\frac{i \alpha  \omega  \text{$\delta$g}_{\text{tx}}}{f^2}
   +\frac{i \alpha  k \text{$\delta$g}_{\text{tt}}}{2 f}
   -\frac{i \alpha  k \text{$\delta $g}_{\text{xx}}}{2 f}
   +\frac{i \alpha  k \text{$\delta$g}_{\text{yy}}}{2 f}\nonumber\\
   +\left(\frac{f'}{f}-\frac{2}{z}\right) \delta \phi _x'
   +\delta \phi _x''&=0\,.
\end{flalign}

\subsection{SSB model}
For the SSB breaking potential $V(X)=X^3$ of section~\ref{sec:breaking2} we have:
\begin{flalign}
   \frac{k \omega  \text{$\delta $g}_{\text{tx}}}{f^2}
   -\frac{\text{$\delta $g}_{\text{tt}} \left(2 z f'-6 f-4 z^4 \left(h'\right)^2+k^2 z^2-4 \alpha ^6 z^6+6\right)}{2 f z^2}\nonumber\\
   +\frac{\text{$\delta $g}_{\text{yy}} \left(-6 f^2+f \left(-z^2 f''+4 z f'+z^4\left(h'\right)^2+k^2 z^2+38 \alpha ^6 z^6+6\right)+\omega ^2 z^2\right)}{2 f^2 z^2}\nonumber\\
   +\frac{\text{$\delta $g}_{\text{xx}} \left(-6 f^2+f \left(-z^2 f''+4 z f'+z^4 \left(h'\right)^2+38 \alpha ^6 z^6+6\right)+\omega ^2 z^2\right)}{2 f^2 z^2}\nonumber\\
   +\left(\frac{3 f'}{2 f}-\frac{2}{z}\right) \text{$\delta $g}_{\text{tt}}'
   +\frac{f' \text{$\delta $g}_{\text{xx}}'}{4 f}
   +\frac{f' \text{$\delta $g}_{\text{yy}}'}{4 f}
   -\frac{3 z^2 h' \text{$\delta $A}_t'}{f}
   -\frac{30 i \alpha ^5 k z^4 \delta \phi_x}{f}
   +\text{$\delta $g}_{\text{tt}}''&=0\,,\label{app:gtt:a3}
\end{flalign}
\begin{flalign}
   \frac{\text{$\delta $g}_{\text{tx}} \left(-2 z f'+6 f+z^4 \left(h'\right)^2-8 \alpha ^6 z^6-6\right)}{f z^2}
   +\frac{k \omega  \text{$\delta $g}_{\text{yy}}}{f}
   -\frac{12 i \alpha ^5 \omega  z^4 \delta \phi _x}{f}\nonumber\\
   +2 z^2 h' \text{$\delta $A}_x'
   -\frac{2 \text{$\delta $g}_{\text{tx}}'}{z}
   +\text{$\delta $g}_{\text{tx}}''&=0\,,
\end{flalign}
\begin{flalign}
   \frac{k \omega  \text{$\delta $g}_{\text{tx}}}{f^2}
   -\frac{\text{$\delta $g}_{\text{tt}} \left(-2 z f'+6 f+2 z^4 \left(h'\right)^2+k^2 z^2+4 \alpha ^6 z^6-6\right)}{2 f z^2}\nonumber\\
   +\frac{\text{$\delta $g}_{\text{yy}} \left(-6 f^2+f \left(-z^2 f''+4 z f'+z^4 \left(h'\right)^2-k^2 z^2+14 \alpha ^6 z^6+6\right)-\omega ^2 z^2\right)}{2 f^2 z^2}\nonumber\\
   +\frac{\text{$\delta $g}_{\text{xx}} \left(6 f^2-f \left(-z^2 f''+4 z f'+z^4 \left(h'\right)^2+26 \alpha ^6 z^6+6\right)+\omega ^2 z^2\right)}{2 f^2 z^2}\nonumber\\
   +\frac{18 i \alpha ^5 k z^4 \delta \phi _x}{f}
   +\left(\frac{3 f'}{4 f}-\frac{2}{z}\right) \text{$\delta $g}_{\text{xx}}'
   -\frac{f' \text{$\delta $g}_{\text{yy}}'}{4 f}
   +\frac{z^2 h' \text{$\delta $A}_t'}{f}
   +\text{$\delta $g}_{\text{xx}}''&=0\,,
\end{flalign}
\begin{flalign}
   -\frac{k \omega  \text{$\delta $g}_{\text{tx}}}{f^2}
   +\frac{\text{$\delta $g}_{\text{tt}} \left(2 z f'-6 f-2 z^4 \left(h'\right)^2+k^2 z^2-4 \alpha ^6 z^6+6\right)}{2 f z^2}\nonumber\\
   +\frac{\text{$\delta $g}_{\text{yy}} \left(6 f^2-f \left(-z^2 f''+4 z f'+z^4 \left(h'\right)^2+k^2 z^2+26 \alpha ^6 z^6+6\right)+\omega ^2 z^2\right)}{2 f^2 z^2}\nonumber\\
   +\frac{\text{$\delta $g}_{\text{xx}} \left(-6 f^2+f \left(-z^2 f''+4 z f'+z^4 \left(h'\right)^2+14 \alpha ^6 z^6+6\right)-\omega ^2 z^2\right)}{2 f^2 z^2}\nonumber\\
   -\frac{6 i \alpha ^5 k z^4 \delta \phi _x}{f}
   -\frac{f' \text{$\delta $g}_{\text{xx}}'}{4 f}
   +\left(\frac{3 f'}{4 f}-\frac{2}{z}\right) \text{$\delta $g}_{\text{yy}}'
   +\frac{z^2 h' \text{$\delta $A}_t'}{f}
   +\text{$\delta $g}_{\text{yy}}''&=0\,,
\end{flalign}
\begin{eqnarray}
   -\frac{k^2 \text{$\delta $A}_t}{f}
   -\frac{k \omega  \text{$\delta $A}_x}{f}
   -\frac{1}{2} h' \text{$\delta $g}_{\text{tt}}'+\frac{1}{2} h' \text{$\delta $g}_{\text{xx}}'
   +\frac{1}{2} h' \text{$\delta $g}_{\text{yy}}'
   +\text{$\delta $A}_t''=0\,,
\end{eqnarray}
\begin{flalign}
   \frac{k \omega  \text{$\delta $A}_t}{f^2}
   +\frac{\omega ^2 \text{$\delta $A}_x}{f^2}
   +\frac{f' \text{$\delta $A}_x'}{f}
   +\frac{h' \text{$\delta $g}_{\text{tx}}'}{f}
   +\text{$\delta $A}_x''&=0\,
\end{flalign}
and
\begin{flalign}
   \frac{\delta \phi _x \left(\omega ^2-3 f k^2\right)}{f^2}
   -\frac{i \alpha  \omega  \text{$\delta $g}_{\text{tx}}}{f^2}
   +\frac{i \alpha  k \text{$\delta $g}_{\text{tt}}}{2 f}
   -\frac{3 i \alpha  k \text{$\delta $g}_{\text{xx}}}{2 f}
   -\frac{i \alpha  k \text{$\delta $g}_{\text{yy}}}{2 f}\nonumber\\
   +\left(\frac{f'}{f}+\frac{2}{z}\right) \delta \phi _x'
   +\delta \phi _x''&=0\,.
\end{flalign}

\section{Numerical techniques}
The numerics are carried out in Mathematica, with the packages xAct~\cite{xAct} and xTras~\cite{Nutma:2013zea}.

The seven background equations of motion provide seven linearly independent solutions. Three are found numerically by imposing in-falling boundary conditions at the horizon, and can be characterized by the starting values of $\delta A_x$, $\delta g_{xx}$ and $\delta\phi_x$ after factoring out their oscillating behaviour. To simplify the numerics, an analytic expansion is made from the horizon to a small offset, before numerically integrating to the boundary, where a similar expansion is made. The remaining four solutions are found analytically as pure gauge solutions.

A mode is characterized as a non-trivial solution to the set of equations of motion, together with a set of boundary conditions. Six of these boundary conditions are  Dirichlet conditions on the metric fluctuations, $\delta A_t$ and $\delta\phi_x$. The last condition is~\eqref{eq:bcs}. As the system is linear, if there is a non-trivial solution to the boundary value problem, it is a linear combination of the seven solutions found above. That means that the sets of boundary values for each of the seven solutions are linearly dependent, and the $7\times7$-matrix formed by them has a vanishing determinant, 
\begin{equation}
\left|\begin{array}{ccccccc}
\delta g_{\text{tt}}(z)_{1} & \delta g_{\text{tx}}(z)_{1} & \delta g_{\text{xx}}(z)_{1} & \delta g_{\text{yy}}(z)_{1} & \delta\phi_x(z)_1 &\delta A_{\text{t}}(z)_{1} &  [\omega^2\delta A_x(z)+\lambda\,\delta A'_x(z)]_{1} \\
\delta g_{\text{tt}}(z)_{2} & \delta g_{\text{tx}}(z)_{2} & \delta g_{\text{xx}}(z)_{2} & \delta g_{\text{yy}}(z)_{2} & \delta\phi_x(z)_2 & \delta A_{\text{t}}(z)_{2} &   [\omega^2\delta A_x(z)+\lambda\,\delta A'_x(z)]_{2} \\
& & & &\cdots
\end{array}\right|_{z\to0}=0\,.
\end{equation}
This provides a convenient way of verifying if a choice of $\omega$ and $k$ lie on a particular mode, up to chosen numerical precision.

\begin{figure}
    \centering
    \includegraphics[width=9cm]{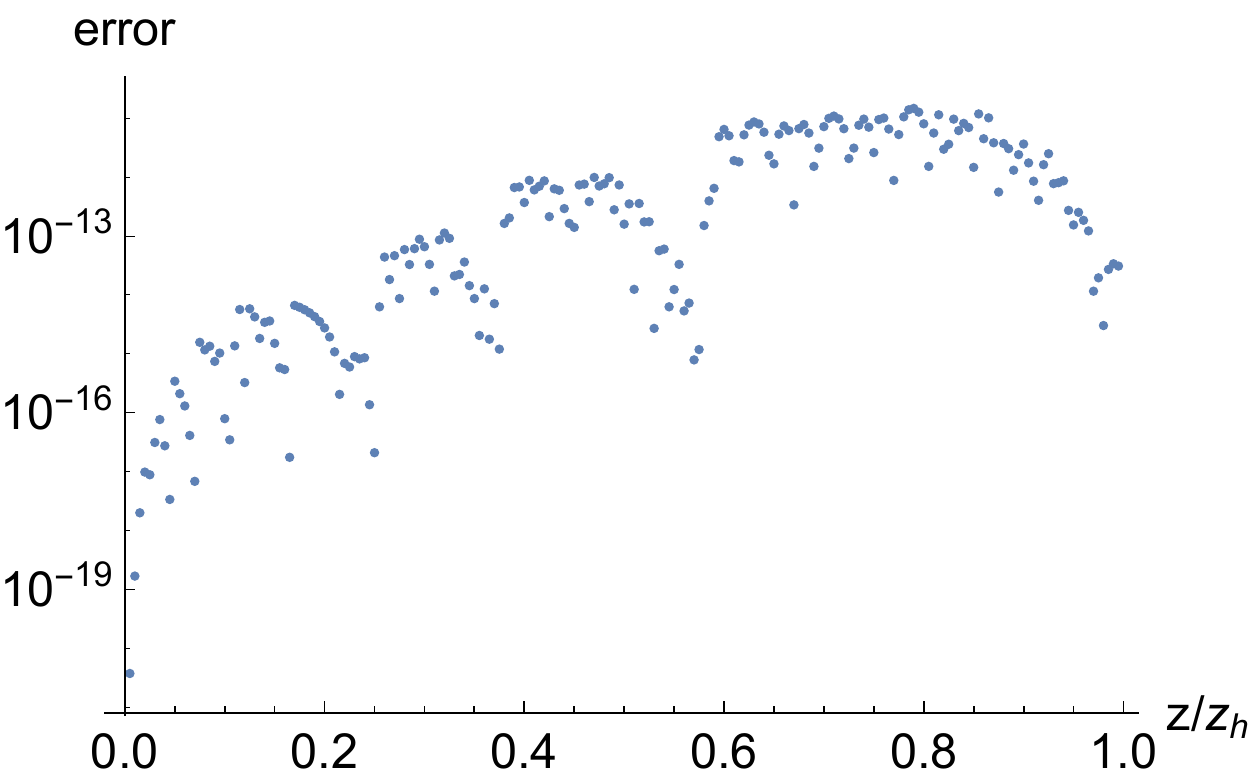}
    \caption{Example of the numerical value accepted for the LHS of \ref{app:gtt:a3} at $\mu/T=4$, $\lambda=1$ and $\alpha^3/T=4$.}
    \label{errorfig}
\end{figure}

The numerical precision is chosen to be overly strong, as to eliminate potential numerical artifacts, an example of the error in the equations of motion is shown in figure \ref{errorfig}.

\label{app2}

\bibliographystyle{JHEP}
\bibliography{plas}
\end{document}